\documentclass[aps,prd,superscriptaddress,nofootinbib,amsmath,amsfonts,preprintnumbers,notitlepage,10pt,english]{revtex4-1}
\usepackage{amsmath}
\usepackage{amssymb}
\usepackage{babel}

\usepackage{bm}
\usepackage{booktabs}
\usepackage{multirow}
\usepackage{array}
\usepackage{enumitem}
\makeatletter

\usepackage{array,multirow,graphicx}
\usepackage{dcolumn}
\usepackage{newlfont}
\usepackage{bm}
\usepackage[colorlinks,citecolor=blue,urlcolor=blue,linkcolor=blue]{hyperref}
\usepackage[figtopcap]{subfigure}
\usepackage{color}

\begin{document}
\date{\today}
\title{Compact stellar model  in higher torsion gravitational theory}
\author{G.G.L. Nashed}
\email{nashed@bue.edu.eg}
\affiliation {Centre for Theoretical Physics, The British University, P.O. Box
43, El Sherouk City, Cairo 11837, Egypt}
\affiliation{Laboratory for Theoretical Cosmology, Tomsk State University of Control Systems and Radioelectronics (TUSUR), 634050 Tomsk, Russia}
\author{Kazuharu Bamba}
\email{bamba@sss.fukushima-u.ac.jp}
\affiliation{Faculty of Symbiotic Systems Science, Fukushima University, Fukushima 960-1296, Japan}

\begin{abstract}
In this study, we address the issue of a spherically symmetrical interior solution to the quadratic form of $f\mathcal{(T)}=\mathcal{T}+\epsilon \mathcal{T}^2$ gravitational theory using a physical tetrad that  provides vanishing components of the off-components of the field equation, in contrast to what exists in the current literature. To be able to formulate the resulting differential equation in a closed form, we employ  the Krori-Barua (KB) ansatz. Using the  KB spacetime form, we derive the analytic form of the energy-density, radial, and tangential pressures and the anisotropic form. All of  these quantities are affected by the dimensional parameter $\epsilon$, which causes them to have a noted  difference from those given in the frame of Einstein general relativity. The derived model of this study exhibits a non-trivial form of torsion scalar, and  it also contains three constants that we drew from the matching of the boundary condition with a line element that  also features a non-trivial form of torsion scalar. Having established the physical conditions that are needed for any real stellar, we check our model and show in detail that it bypasses all of these. Finally, we analyze the model's stability utilizing  the Tolman-Oppenheimer-Volkoff equation and  adiabatic index and show that our model satisfies these.
\end{abstract}


\maketitle
\section{Introduction}
Shortly after the development of general relativity (GR), different gravitational theories were still being followed. Among these was one  given by H. Weyl that sought  to unify gravitation with electromagnetism in (1918) \cite{1918SPAW.......465W}; an endeavor that was unsuccessful. Einstein in (1928)  \cite{doi:10.1002/3527608958.ch37}  tried to achieve the same goal as Weyl and he drew on the geometry of  Weitzenb\"ock. In this geometry, one must  to introduce a tetrad field as the dynamical variable in the theory, unlike GR, whose dynamical variable is the metric. The tetrad field consists  of 16 components, which means that Einstein thought that the additional six components from GR could prescribe the parameters of the electromagnetic field. Nonetheless, researchers  showed that these extra six components were linked to  local Lorentz invariance in the theory \cite{MuellerHoissen:1983vc}.

Despite the failure   of Weyl  and Einstein's efforts  they still  supplied  the notion of gauge theories, and so the search  for a gauge theory of gravitation began in earnest \cite{ORaifeartaigh:1997dvq,Blagojevic:2013xpa}.
In (1979),  Hayashi and  Shirafuji \cite{PhysRevD.19.3524} proposed a gravitational theory that they termed  ``new GR'' that was a gauge theory of the translation group. This theory entailed three free parameters to be experimentally-determined.

Another theory that was built on  Weitzenb\"ock geometry is the teleparallel equivalent of GR (TEGR). The theories of TEGR and GR are commensurate with  the field equations  however, at the level of actions, they differ in their  total divergence  term \cite{Maluf_2013,Wu_2012}. Despite this  conceptual  difference,  TEGR and GR  do not  differ   experimentally. In the TEGR theory, gravity is encoded in the torsion, with a  vanishing curvature, unlike GR in which gravity is encoded in the metric with vanishing torsion \cite{Li:2010cg,Shirafuji:1997wy,Krssak:2015lba,Nashed:2003ee,Bahamonde:2015zma}.

 Compact objects are formed when stars  exhaust all of their nuclear fuel. Nowadays, such compact stars are of interest to scientists, especially the class of neutron stars that are boosted by neutron degeneracy pressure against the attraction of gravity. Meanwhile another type of compact star exists namely  the white dwarf which is boosted  by electron degeneracy pressures against  gravity. It is well known that the first precise vacuum solution of GR is that by Schwarzschild  \cite{Schwarzschild:1916uq}.  Thereafter, many solutions to investigate static  compact stellar formulations  were derived. The search for interior GR solutions that describe  realistic compact stellar objects became a fertile  domain  to scientists. Einsteinian GR became a highly  important theory that influenced the study of  compact
stellar objects following the discovery of quasars in  (1960). Using the equation of state (EoS), one can analyze the stability structure of a compact stellar
object. By means of EoS, one can also investigate the physical stellar behavior using the Tolman-Oppenheimer-Volkov (TOV) equation, which is the general relativistic equation of  stellar constructions.  

In order to study stellar configurations in GR  the distribution of matter must be assumed to be isotropic, which means that the radial and tangential pressures are equal. However, in reality, such an assumption is not held and one can find that the two components of radial and tangential pressures are not equal and that the difference between them  results in the anisotropy. Such stellar configurations exhibit unequal radial and tangential pressures. In  1933 Leimatre  was the first to predict anisotropic model \cite{Lemaitre:1933gd}. Moreover, researchers showed that to reach the maximum on the surface of a star then, the radial pressure decreased  and ultimately vanished at the center \cite{PhysRevD.70.024010,Mak:2003kw}.  Many factors must be taken into account in assuming that a star is anisotropic; among these is a high-density regime in which nuclear interactions are relativistically treated \cite{Ruderman:1972aj,1975ARA&A..13..335C}.  Furthermore, the existence of a solid core or 3A-type  superfluid may cause a star to become anisotropic \cite{Kippenhahn:1493272}. Another  source exists that causes stellar anisotropy namely,  the strong magnetic field \cite{article3}. The slow rotation can be considered a source of anisotropy \cite{Herrera:1995pm}. Letelier  showed that the combination between perfect and null fluids can be prescribed as anisotropic in nature  \cite{PhysRevD.22.807}.  Many reasons can be taken into account that can yield the anisotropic-like pion condensation \cite{PhysRevLett.29.382}, including the strong electromagnetic field \cite{PhysRevD.70.067301} and phase transition \cite{1980JETP...52..575S}.  Dev and Gleiser \cite{Dev2003,Dev2002} and Gleiser and Dev \cite{GLEISER_2004}  investigated   operators that affect the pressure to be anisotropic. Researchers showed that the effect of  shear, electromagnetic field,
etc $\cdots$  on self-bound systems can be neglected  if the system
is   anisotropic \cite{Ivanov_2010}. Systems that consists of scalar fields such as  boson stars possess anisotropic qualities \cite{Schunck:2003kk}. Gravastars and wormholes are considered anisotropic models \cite{Morris:1988cz,Cattoen:2005he,DeBenedictis_2006}. An application of an  anisotropic model to the stable configuration of neutron stars has also been discussed \cite{Bowers:1974tgi} and they show that  anisotropy
could exhibit non-negligible effects on  equilibrium mass and surface red-shift. A fine study that describes the origin and influence of anisotropy can be found in \cite{Chan:2002bn,10.1093/mnras/287.1.161}. Super dense  anisotropic neutron stars have also been considered, and the conclusion that  no limiting mass of such stars exits was discussed \cite{1975A&A....38...51H}. The issue of the stability of such stars was analyzed in \cite{etde_7103699} and a conclusion was drawn that such stability is  similar to that of isotropic stars.  Many anisotropic models deal with  anisotropic pressure in the energy-momentum  tensor in the compositions of matter. Several exactly spherically symmetrical solutions of interior stellar  have also been developed \cite{PhysRevD.26.1262,1984CaJPh..62..239K,1993MNRAS.262.1088B,1999MNRAS.302..337B,Hanafy:2015yya,1993Ap&SS.201..191B,Barreto2007,Coley_1994,1994MNRAS.271..463M,Singh1995,
Hern_ndez_1999,Harko20,Nashed:2018nll,Shirafuji:1997wy,Awad:2017sau,
1995AuJPh..48..635P,PhysRevLett.92.051101,Bohmer2006,Nashed:2011fg,Boehmer2007az,Esculpi2007,Shirafuji:1995xc,Khadekar7,Karmakar2007,Abreu2007,
IVANOV2010,PhysRevD.79.087505,Nashed:2008ys,Mak2003,doi10.1142/S0217732302008149,Maharaj1989,Chaisi2006,PhysRevD.77.027502,Chaisi2005,Gokhroo1994,PhysRevD.80.064039,
Thomas12,2007IJMPD..16.1479T,Elizalde:2020icc,Tikekar2005,Thirukkanesh2008,Finch1989,Sharma:2013lqa,2015Ap&SS.356..285P,Bhar2015}.

This paper is structured as  follows: In Section \ref{S2}, we present the ingredient  bases of  TEGR theory. In Section \ref{S3}, we apply the non-vacuum charged field equations of $f\mathcal{(T)}$ theory to a non-diagonal tetrad, and we derive the non-vanishing components of the differential equations. Then we show that the number of the nonlinear differential equations is less than the number of unknowns, and, therefore, we make  two different assumptions, one related to the $g_{rr}$ potential metric and the other to the anisotropy,   and derive a new solution in Section \ref{S3}. In Section \ref{S4}, we use the matching condition to match the constants of integration involving in the interior solution with an exterior spherically symmetric spacetime that its torsion scalar depends on the radial coordinate. In Section \ref{S5}, we discuss the physical content of the interior  solution and show that it  possesses many merits that make it physically acceptable. Among these is  that it satisfies the energy and causality conditions,. In Section \ref{S6}, we study the stability using the TOV equation and the adiabatic index and show that  our interior solution exhibits static stability. In Section  \ref{S7} we conclude and discuss the main results of this study.
\section{Ingredient tools of $f(\mathcal{T})$ gravitational theories}\label{S2}
In this section, we briefly present  the $f(\mathcal{T})$ gravitational theory. In torsional
formulations of gravity using the tetrad fields $b^{i}{_{\mu}}$ as   dynamical variables proves to be convenient,
which form an orthonormal basis for  tangent space at each point  of spacetime.
These are related to the metric through the following:
\begin{equation}\label{q3}
 {\textit g_{\mu \nu} :=  \eta_{i j} {b^i}_\mu {b^j}_\nu,}
\end{equation}
 with $\eta_{i j}=(-,+,+,+)$ being the  $4$-dimensional Minkowskian metric of
the tangent space (The Greek indices are used for the coordinate space and the Latin
indices for the tangent one).  The curvatureless Weitzenb\"{o}ck
connection   is  introduced as  $\overset{{\textbf {w}}}{\Gamma}^\lambda_{\mu \nu} := {b_i}^\lambda~
\partial_\nu b^{i}_\mu$  \cite{Wr}, and thus the torsion
tensor is defined as the following:
\begin{equation}
\label{torsten}
{\mathcal{T}^\alpha}_{\mu \nu} :=
\overset{{\textbf {w}}}{\Gamma}^\alpha_{ \nu\mu }-\overset{{\textbf {w}}}{\Gamma}^\alpha_{\mu \nu}
={b_i}^\alpha
\left(\partial_\mu{b^i}_\nu-\partial_\nu{b^i}_\mu\right),
\end{equation}
which carries all of the information on the gravitational field. Finally, in contracting the
torsion tensor, we obtain the torsion scalar as the following:
\begin{equation}
\label{Tor_sc}
\mathcal{T}\equiv\frac{1}{4}
\mathcal{T}^{\rho \mu \nu}
\mathcal{T}_{\rho \mu \nu}
+\frac{1}{2}\mathcal{T}^{\rho \mu \nu }\mathcal{T}_{\nu \mu\rho }
-\mathcal{T}_{\rho \mu }^{\ \ \rho }\mathcal{T}_{\
\ \ \nu }^{\nu \mu }.
\end{equation}
 When $\mathcal{T}$ is used   as the Lagrangian in the action of teleparallel gravity the obtained
theory is the TEGR, as variation with respect to the tetrads lead  to
exactly the same equations using GR.

Inspired by the $f(R)$ extensions of GR, one can extend $\mathcal{T}$ to $f\mathcal{(T)}$,
obtaining  $f\mathcal{(T)}$ gravity, as determined by the following action \cite{Cai:2015emx}:
\begin{equation}\label{q7a}
{\cal L}=\frac{1}{2\kappa}\int |b|f\mathcal{(T)}~d^{4}x,
\end{equation}
where $|b|=\sqrt{-g}=\det\left({b^a}_\mu\right)$ is the determinant of the metric and
$\kappa$  is  a dimensional constant defined as $\kappa =8\pi$.

In this work, we seek to study an interior solution within the framework of
$f(\mathcal{T})$ gravity. Hence, in action (\ref{q7a}) we add the Lagrangian  matter  too.
Therefore, the considered action  is the following:
\begin{equation}\label{q7}
{\cal L}=\frac{1}{2\kappa}\int |b|f\mathcal{(T)}~d^{4}x+\int |b|{\cal L}_{matter}~d^{4}x,
\end{equation}
where  ${\cal L}_{matter}$, is the  Lagrangian matter.

Variation of action (\ref{q7}) with respect to the tetrads  leads to
\cite{Cai:2015emx}:
\begin{eqnarray}\label{q8a}
& & {\mathop{\mathcal{ E}}}^\nu{}_\mu={S_\mu}^{\rho \nu} \partial_{\rho} \mathcal{T}
f_{\mathcal{T}\mathcal{T}}+\left[b^{-1}{b^i}_\mu\partial_\rho\left(b{b_i}^\alpha
{S_\alpha}^{\rho \nu}\right)-{\mathcal{T}^\alpha}_{\lambda \mu}{S_\alpha}^{\nu \lambda}\right]f_\mathcal{T}
-\frac{f}{4}\delta^\nu_\mu +\frac{1}{2}\kappa{{{\mathfrak{
T}}^{{}^{{}^{^{}{\!\!\!\!\scriptstyle{matter}}}}}}}^\nu_\mu \equiv0,
\end{eqnarray}
with $f := f(\mathcal{T})$, \ \   $f_{\mathcal{T}}:=\frac{\partial f(\mathcal{T})}{\partial \mathcal{T}}$, \ \
$f_{\mathcal{T}\mathcal{T}}:=\frac{\partial^2
f\mathcal{(T)}}{\partial \mathcal{T}^2}$. Here     ${{{\mathfrak
T}^{{}^{{}^{^{}{\!\!\!\!\scriptstyle{matter}}}}}}}^\nu_\mu$  is  the
energy-momentum tensor  of an anisotropic fluid defined as the following:
\begin{equation}
{{{\mathfrak
T}^{{}^{{}^{^{}{\!\!\!\!\scriptstyle{matter}}}}}}}^\mu_\nu=\left(\frac{p_t}{c^2}+\rho\right)u^\mu u_a+p_t\delta_a{}^\mu+(p_r-p_t)\xi_a \xi^\mu\,,
\end{equation}
where $c$ is the speed of light. Here, $u_\mu$  the time-like vector is defined as $u^\mu=[1,0,0,0]$ and $\xi_\mu$ is a unit space-like vector in the radial direction, defined as $\xi^\mu=[0,1,0,0]$ such that $u^\mu u_\mu=-1$ and $\xi^\mu\xi_\mu=1$. In this paper, $\rho$ is the energy-density and $p_r$  and $p_t$ are the radial and    tangential pressures, respectively.
The above equations determine $f(\mathcal{T})$ gravity in 4-dimension space.

\section{Stellar  equations in f(R) gravitational theory}\label{S3}
 We are using  the following spherically symmetrical spacetime, which demonstrates two unknown functions as\footnote{We do not discuss the issue of spin connection in the frame of $f(T)$ gravity  because its value is vanishing identically for the tetrad (\ref{tet1}) \cite{DeBenedictis:2016aze,Ilijic:2018ulf,Bahamonde_2019} which we use as the  main input in this study.}:
\begin{eqnarray} \label{met12}
& &  ds^2=-e^{\alpha(r)}dt^2+e^{\beta(r)}dr^2+r^2d\Omega\,, \qquad {\textrm where} \qquad d\Omega=(d\theta^2+\sin^2d\phi^2)\,,  \end{eqnarray}
where $\alpha(r)$ and $\beta(r)$ are unknown functions.   The above metric of Eq. (\ref{met12}) can be reproduced from the following covariant tetrad field ~\cite{Bahamonde_2019}
\begin{equation}
b^a{}_{\mu}=\left(
\begin{array}{cccc}
e^{\alpha(r)/2} & 0 & 0 & 0 \\
0 & e^{\beta(r)/2} \cos (\phi ) \sin (\theta ) & r \cos (\phi ) \cos (\theta )  & -r \sin (\phi ) \sin (\theta )  \\
0 & e^{\beta(r)/2} \sin (\phi ) \sin (\theta )  & r \sin (\phi ) \cos (\theta )  & r \cos (\phi ) \sin (\theta ) \\
0 & e^{\beta(r)/2} \cos (\theta ) & -r \sin (\theta ) & 0 \\
\end{array}
\right)\label{tet1}\,.
\end{equation}

 The torsion scalar of Eq. (\ref{tet1}) takes the form:
\begin{eqnarray} \label{Tor}
  {\mathcal T(r)}=\frac{2(1-e^{\beta/2})(e^{\beta/2}-1-r\alpha')}{r^2e^{\beta}}\,,
  \end{eqnarray}
 where $\alpha\equiv \alpha(r)$, $\beta\equiv \beta(r)$, $\alpha'=\frac{d\alpha}{dr}$, $\alpha''=\frac{d^2\alpha}{dr^2}$ and $\beta'=\frac{d\beta}{dr}$.  In this study, we consider the model of as
 \begin{eqnarray} \label{qud}
f(\mathcal{T})=\mathcal{T}+1/2\epsilon \mathcal{T}^2\,,
\end{eqnarray}
 where $\epsilon$ is a dimensional  parameter which has a unit of distance squared \cite{,Bahamonde_2019}.
 For the tetrad (\ref{tet1}), the non-vanishing components of the field equations
 (\ref{q8a}) when $f(\mathcal{T})=\mathcal{T}+1/2\epsilon \mathcal{T}^2$  take the following form \cite{Cai:2015emx}:
 \begin{eqnarray} \label{fesxx}
&&\frac{8\pi G}{c^2} \rho=\frac{e^{-2\beta}}{r^4}\Big[2\epsilon\,e^{\beta/2}\{2r\beta'[2r\alpha'+3]-4r^2\alpha''-r^2\alpha'^2+8\}+8\epsilon e^{3\beta/2}+4\epsilon r^2\alpha''(1+e^{\beta})+e^{2\beta}(r^2-\epsilon)+r\beta'\Big\{e^{\beta}(r^2-6\epsilon)\nonumber\\
&&-2r\epsilon(3+e^{\beta}) \alpha'-6\epsilon\Big\}+\epsilon r^2 \alpha'^2(1+e^{\beta})-e^{\beta}(r^2+18\epsilon)-5\epsilon\Big]\,,\nonumber\\
&& \frac{8\pi G}{c^4} p_r=\frac{e^{-2\beta}}{r^4}\Big[\epsilon\,e^{\beta/2}\{12r\alpha'+4 r^2\alpha'^2+8\}+e^{2\beta}\{\epsilon-r^2\}-r^2\epsilon\,\alpha'^2\{3+e^{\beta}\}+r\,\alpha'\{(r^2-6\epsilon)e^{\beta}-6\epsilon\}+e^{\beta}(r^2-6\epsilon)
-3\epsilon\Big]\,,\nonumber\\
&&\frac{8\pi G}{c^4} p_t=\frac{e^{-2\beta}}{4r^4}\Big[2 \epsilon e^{\beta/2}\{4r^2(3+r\alpha')\alpha''+r^3\alpha'^3-2\alpha'^2\,r^2(r\beta'-3)-12r^2\alpha'\beta'-12r\beta'-16\}+8r\epsilon \alpha'e^{3/2\beta}\nonumber\\
&&+2r^2\alpha''(e^{\beta}[r^2-6\epsilon]-4\epsilon r\alpha'-6\epsilon)-4\epsilon e^{2\beta}-2\epsilon r^3\alpha'^3+r^2\alpha'^2(6\epsilon r \beta'+e^{\beta}(r^2-6\epsilon)-6\epsilon)-r\alpha'\Big([e^{\beta}(r^2-6\epsilon)-18\epsilon]r\beta'\nonumber\\
&&+2e^{\beta}(6\epsilon-r^2)-4\epsilon\Big)-2r\beta'[e^{\beta}
(r^2-6\epsilon)-6\epsilon]+12\epsilon(e^{\beta} +1) \Big]\,,
\end{eqnarray}
where we use the form of  anisotropic energy momentum-tensor that exhibits the form $T_\mu{}^\nu=[\rho\, c^2, \,p_r,\, p_t, and \, \,\,p_t]$.  It is noteworthy  to mention that if the dimensional parameter is $\epsilon=0$ and we will return to the differential equations given in \cite{Roupas:2020mvs,2020arXiv200211471N}.
We define the anisotropic parameter of the stellar as the following:
\begin{eqnarray} \label{anis}
&&\Delta(r)=\frac{8\pi G}{c^4}\Big[p_{_{_t}}-p_r\Big]=\frac{e^{-2\beta}}{4r^4}\Big[2 \epsilon e^{\beta/2}\{4r^2(3+r\alpha')\alpha''+r^3\alpha'^3-2\alpha'^2\,r^2(r\beta'+1)-12r\alpha'[2+r\beta']-4[3r\beta'+8]\}\nonumber\\
&&+8r\epsilon \alpha'e^{3/2\beta}+2r^2\alpha''(e^{\beta}[r^2-6\epsilon]-4\epsilon r\alpha'-6\epsilon)+4(r^2-2\epsilon) e^{2\beta}-2\epsilon r^3\alpha'^3+r^2\alpha'^2(6\epsilon r \beta'+e^{\beta}(r^2-2\epsilon)+6\epsilon)\nonumber\\
&&-r\alpha'\Big([e^{\beta}(r^2-6\epsilon)-18\epsilon]r\beta'+2e^{\beta}(r^2-6\epsilon)-28\epsilon\Big)-2r\beta'[e^{\beta}
(r^2-6\epsilon)-6\epsilon]+24\epsilon(2e^{\beta} +1)-4r^2 e^{\beta}\Big]\,.
\end{eqnarray}

Moreover, we introduce the characteristic density thus
\begin{equation}\label{eq:rho_star}
\rho_\star \equiv \frac{c^2}{4\pi G R^4}\,,
\end{equation}
which we will apply to re-scale the density, radial and tangential pressures, to obtain the dimensionless variables:
\begin{equation}\label{eq:rho+p_dless}
\tilde{\rho} = \frac{\rho}{\rho_\star}\,,\; \qquad
\tilde{p}_r = \frac{p}{\rho_\star c^2}\,,\;\qquad
\tilde{p}_t = \frac{p_t}{\rho_\star c^2}\,,\;\qquad
\tilde{\Delta} = \frac{\Delta}{\rho_\star c^2}\,,\;
\end{equation}
 Moreover,  we re-scale the dimension parameter $\epsilon$  to put it in a dimensionless form as:
\begin{equation}\label{eq:rho+p_dless1}
\varepsilon=\frac{\epsilon}{\chi^2R^2}\,,\;
\end{equation}
where  $\chi$ is a dimensionless variable defined as
\begin{eqnarray} \label{dim}
\chi\equiv \frac{r}{R} \in[0, 1]\,,
\end{eqnarray}
 where $R$ is the radius of the stellar.

Equations (\ref{fesxx}) are three nonlinear differentials in five unknowns, $\alpha$, $\beta$, $\rho$, $p_r$ ,and $p_{_{_t}}$.  Therefore, to adjust the unknowns with the differential equations, we require two additional conditions. These two constrains are the  Krori-Barua (KB) ansatz which takes the following form:
\begin{align}\label{eq:pot}
\alpha(\chi) =f_0 \chi^2+f_1\,,\quad
\beta(\chi) =f_2 \chi^2\,.
\end{align}
The parameters $f_0$, $f_1$, $f_2$ in the  ansatz (\ref{eq:pot}) are dimensionless and their value will be determined from the matching conditions on the boundary. The KB (\ref{eq:pot}) is a physical ansatz  as the gravitational potentials and their derivatives   at the center are finite.

  Using Eq. (\ref{eq:pot})  in Eq. (\ref{fesxx}) we obtain the following:
 \begin{eqnarray} \label{sol}
 && \tilde{\rho}(\chi)=\frac{ e^{-2f_2\chi^2}}{\chi^4}
 \Big[8\varepsilon e^{f_2\chi^2/2}[2-f_0\chi^4(f_0-4f_2)-\chi^2(2f_0-3f_2)]+ 8\varepsilon e^{3f_2\chi^2/2}+e^{2f_2\chi^2}(\chi^2-\varepsilon)+e^{f_2\chi^2}\Big[2\varepsilon\Big(2\chi^2\{2f_0-3f_2\}\nonumber\\
 &&+2f_0\chi^4(f_0-2f_2)-9\Big)+(2f_2\chi^2-1)\chi^2\Big]+\varepsilon\{4\chi^2(2f_0-3f_2)+4\chi^4f_0(f_0-6f_2)-5\}\Big]\,,\nonumber\\
 &&\tilde{p}_r(\chi)=\frac{e^{-2f_2\chi^2}}{\chi^4}\Big[8\varepsilon e^{f_2\chi^2/2}(1+f_0\chi^2[3+2f_0\chi^2])+e^{2f_2\chi^2}(\varepsilon-\chi^2)+e^{f_2\chi^2}(2f_0\chi^4+\chi^2-2\varepsilon[3+2f_0\chi^2\{3+\chi^2\}])\nonumber\\
 &&
 -3\varepsilon(2f_0\chi^2+1)^2\Big]\,,\nonumber\\
 && \tilde{p}_{_{_t}}(\chi)=\frac{e^{-2f_2\chi^2}}{\chi^4}\Big[4\varepsilon e^{f_2\chi^2/2}[3(f_0-f_2)\chi^2+f_0(5f_0-6f_2)\chi^4+f_0{}^2(f_0-2f_2)\chi^6-2]+4e^{3f_2\chi^2/2}\varepsilon f_0 \chi^2-\,\varepsilon e^{2f_2\chi^2}\nonumber\\
 &&+ e^{f_2\chi^2}[f_0\chi^6(f_0-f_2)+\chi^4(2f_0-f_2-6\varepsilon f_0[f_0-f_2])-6\varepsilon \chi^2(2f_0-f_2)+6\varepsilon]+\varepsilon[3-4f_0{}^2\chi^6(f_0-3f_2)-2f_0(7f_0-9f_2)\chi^4\nonumber\\
 &&-2(2f_0-3f_2)\chi^2]\Big]\,,\nonumber\\
 && \Delta(\chi)=\frac{e^{-2f_2\chi^2}}{\chi^4}\Big[4\varepsilon e^{f_2\chi^2/2}[f_0{}^2(f_0-2f_2)\chi^6+f_0(f_0-6f_2)\chi^4-3(f_0-f_2)\chi^2-4]+4\varepsilon f_0\chi^2e^{3f_2\chi^2/2} +(\chi^2-2\varepsilon)e^{2f_2\chi^2}\nonumber\\
 &&+e^{f_2\chi^2}\Big(f_0[f_0-f_2]\chi^6-\chi^4[f_2+2f_0\varepsilon(f_0-3f_2)]-\chi^2(1-6\varepsilon f_2)+6\varepsilon\Big)+2\varepsilon(1+f_0\chi^2)(3-2f_0\chi^4[f_0-3f_2]+\chi^2[3f_2+f_0])\Big]\,.\nonumber\\
 &&
  \end{eqnarray}
  Using Eq. (\ref{eq:pot}) in (\ref{Tor}) we get
  \begin{eqnarray}\label{Torv}
 &&{\mathcal T(\chi)}=\frac{2e^{-f_2\chi^2}[1-e^{f_2\chi^2/2}+2f_0\chi^2][e^{f_2\chi^2/2}-1]}{\chi^2}\,,
  \end{eqnarray}
  which shows that the torsion scalar has a non-trivial value and its leading order is $O\left(\frac{1}{\chi^2}\right)$.
      \section{\bf  Matching conditions }\label{S4}
   Due to the fact that  solution (\ref{sol}) exhibits a non-trivial torsion scalar, we should therefore match it with an spherically symmetrical exterior solution that exhibits a non-constant torsion scalar.   For this goal, we are going to match solution (\ref{sol}) with the neutral one presented in \cite{Bahamonde_2019}. This  spherically symmetrical solution is presented in \cite{Bahamonde_2019} takes the following form:
    \begin{eqnarray}\label{Eq1} && ds^2= -A(\chi)dt^2+B(\chi)dr^2+r^2d\Omega_2{}^2, \quad A(\chi)=\mu^2+\varepsilon\frac{13-99\mu^2+128\mu^3-45\mu^4+3\mu^6+12(1-3\mu^2)\ln(\mu)}{6\chi^2(1-\mu^2)^2}, \nonumber\\
 && B(\chi)=\mu^{-2}-\varepsilon\frac{1+24\mu-12\mu^2-64\mu^3+75\mu^4-24\mu^5+12\ln(\mu)}{6\chi^2(1-\mu^2)}, \quad \textrm{where} \quad \mu=\sqrt{1-\frac{2u}{\chi}}\,,
 \end{eqnarray}
where $2u<\chi$ and $u=M/R$ { and $d\Omega_2{}^2$ is the line element on the unit 2-
sphere}. We must match the interior spacetime metric (\ref{met12}) with
the exterior one given by Eq. (\ref{Eq1}) at the
boundary of the star, $r =R$ in addition to the vanishing of the radial pressure at the surface of the star. These boundary conditions yield the following:
\begin{eqnarray}\label{Eq2} \alpha(r=R)=\ln\,A, \qquad \qquad \beta(r=R)=ln\, B\,, \qquad \qquad  \tilde{p}_r=0.
 \end{eqnarray}
  { One  the basis of the above conditions, we obtain a complicated form of the parameters $f_0$, $f_1$ and $f_2$. The form that these parameters take is the following numerical one when $\varepsilon=0.5$ for the white
dwarf companion pulsar $PRS J0437-4715$ whose  mass is  $1.44^{+0.07}_{-0.07}M_\odot$, where $1.44$ is the mean value and $\pm0.07$ is the error and its radius  $13.6^{+0.9}_{-0.8}$ km, respectively}\footnote{{ Here the unit of the mass of the stellar  is the mass sun and the unit of the radius  is a Kilometer.}}:
  \begin{eqnarray}\label{Eq3} f_0=0.2232587486, \qquad \qquad f_1=-0.5918023605
,\qquad \qquad f_2=.3685944011\,.\end{eqnarray} We will use the above values of  parameters $f_0$, $f_1$ and $f_2$ for  the reminder of this study.
\section{\bf   Terms of physical  viability of the solution (\ref{sol}) }\label{S5}
Now, we are going to check solution (\ref{sol}) to determine if it  describes a real physical stellar space. To do this, we are going to state the following necessary  conditions which are important for any actual star.
\subsection{\bf The energy--momentum tensor}
\begin{figure}
\centering
\subfigure[~The density of solution  (\ref{sol})]{\label{fig:dnesity}\includegraphics[scale=0.3]{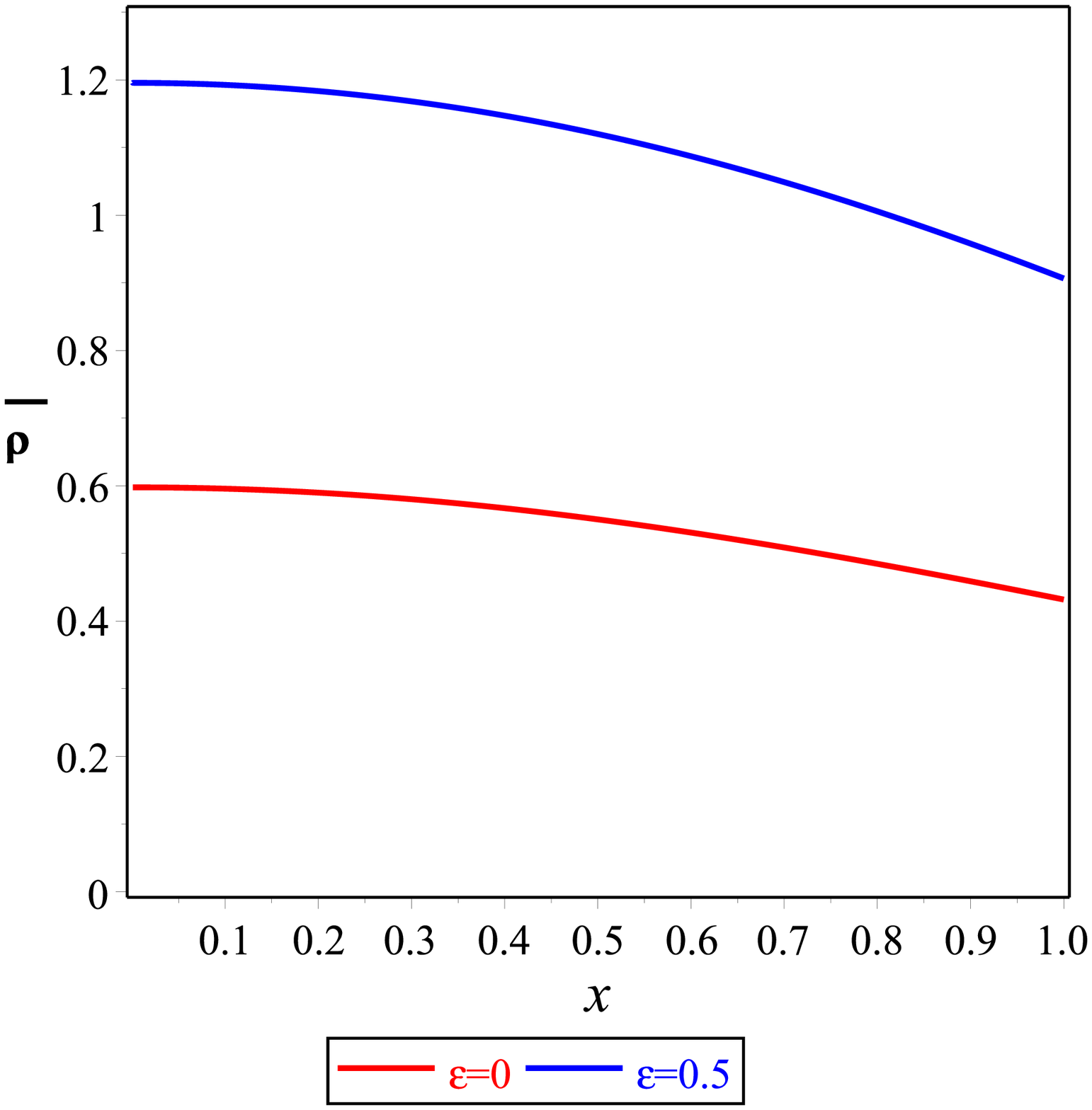}}
\subfigure[~The radial pressure of solution  (\ref{sol})]{\label{fig:pressure}\includegraphics[scale=.3]{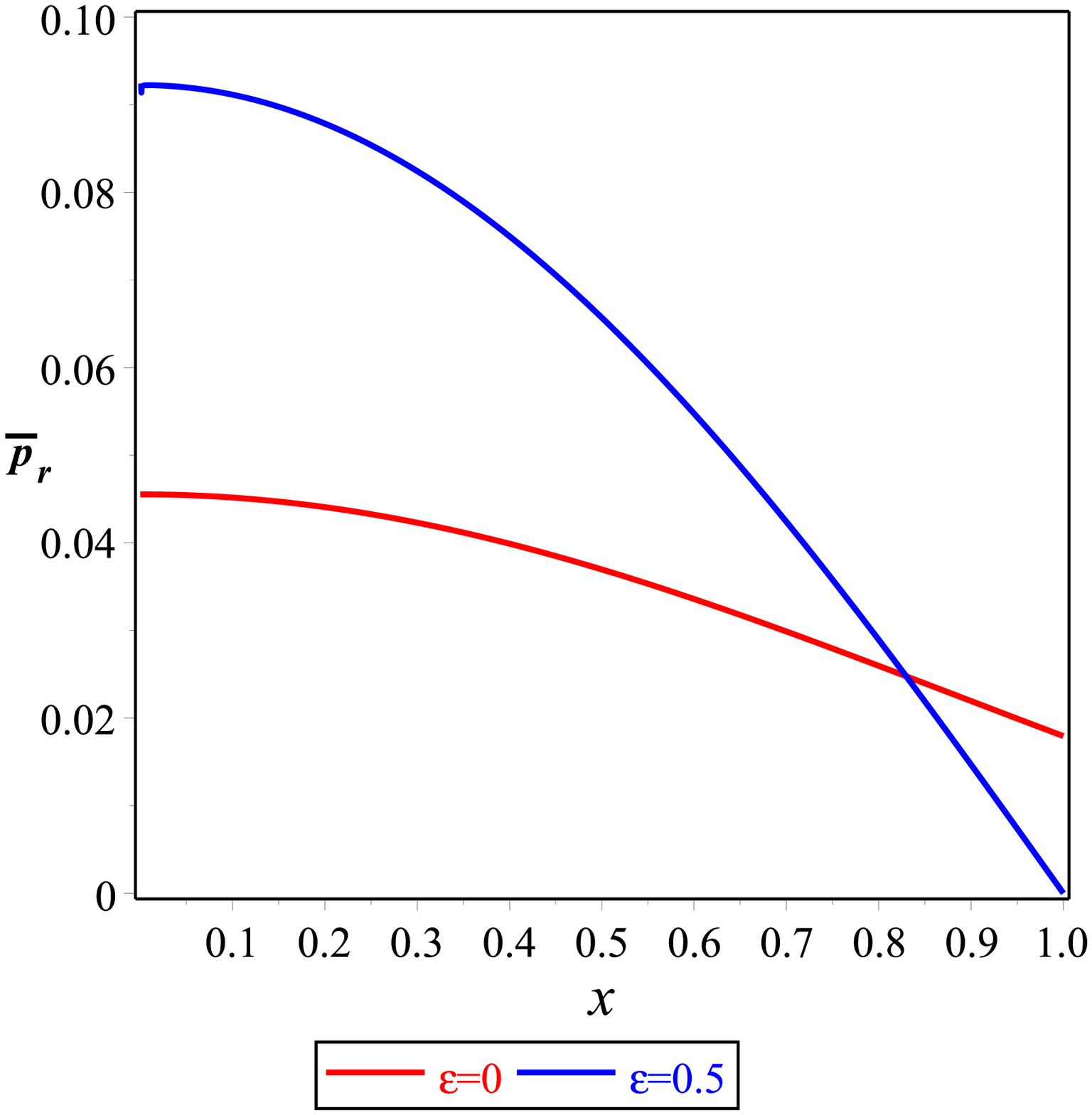}}
\subfigure[~The transverse pressure of solution  (\ref{sol})]{\label{fig:pressuret}\includegraphics[scale=.3]{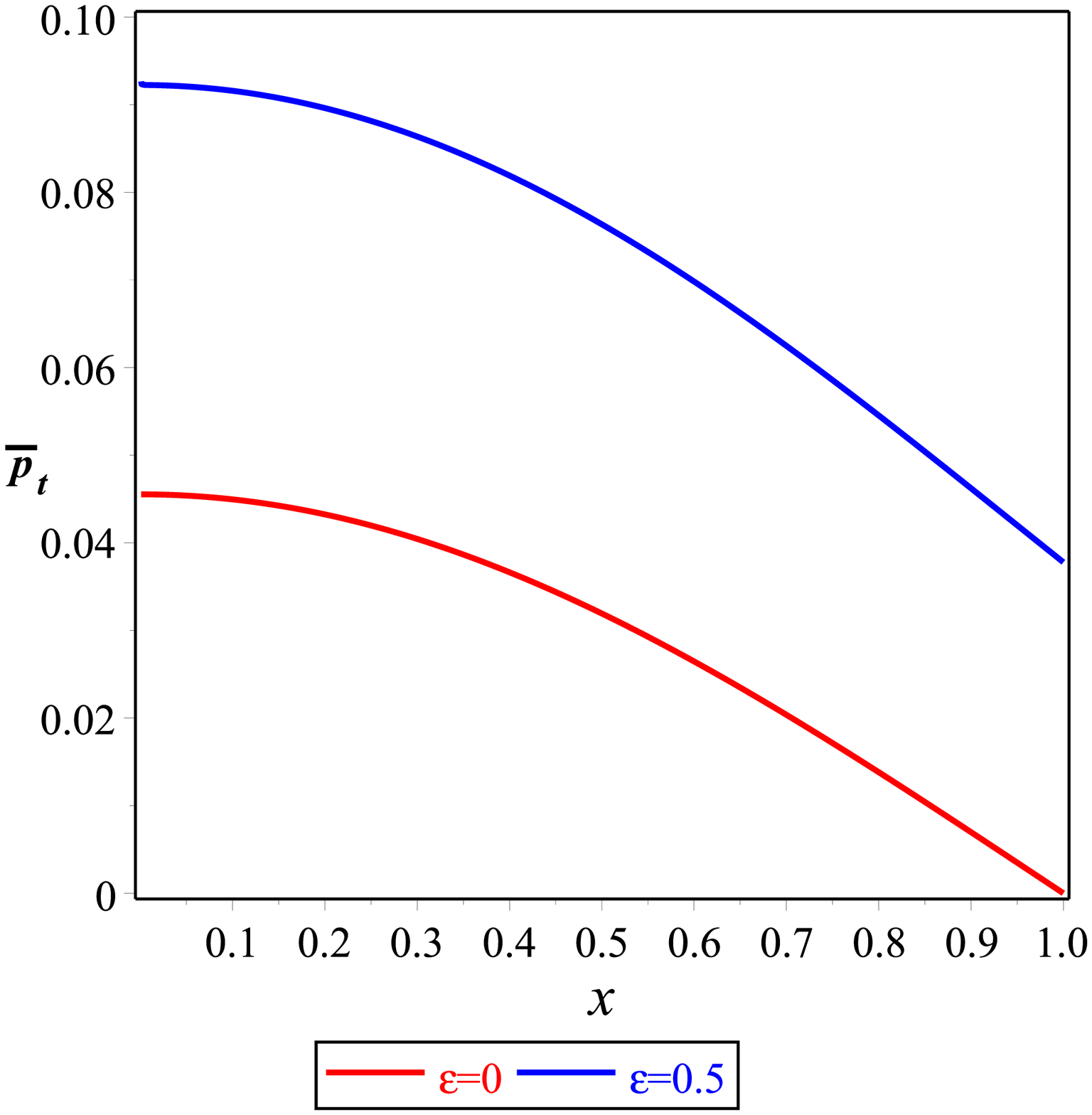}}
\caption[figtopcap]{\small{{Density, radial and transverse pressures  when $\varepsilon=0$ and $\varepsilon=0.5$.}}}
\label{Fig:1}
\end{figure}

 It is known that for a true  interior solution the components of the energy-momentum tensor, energy-density, radial and transverse  pressures must demonstrate positive values. Moreover, the components of the energy-momentum  tensor must be finite  at the center of the star after that they decrease    toward  the surface of the stellar space and the radial pressure must be greater than  the tangential one. Figure \ref{Fig:1} indicates the behavior of the energy-momentum components, density, radial and transverse pressures. It is clear from this figure that $\rho(\chi=0)_{\varepsilon=0}=0.5976904312$, $\tilde{\rho}(\chi=0)_{\varepsilon=0.5}=1.195668990$, $\tilde{p}_r(\chi=0)_{\varepsilon=0}=\tilde{p}_t(\chi=0)_{\varepsilon=0}=0.04553459825$, and $p_r(\chi=0)_{\varepsilon=0.5}=p_t(\chi=0)_{\varepsilon=0.5}=0.0922628584$. Also, Fig.  \ref{Fig:1}, shows that the density, radial and tangential pressures are decreasing toward the stellar surface. Moreover  Fig. \ref{Fig:1} shows that the values of  the components of the energy momentum  at the center in case $\varepsilon=0$ are smaller than those when  $\varepsilon \neq0$. We can also deduce from  Fig. \ref{Fig:1}  that the density, radial and  tangential pressures move to the surface of the stellar more rapidly in case $\varepsilon=0$  than in the case of $\varepsilon=0.5$.  Finally,  it  can be seen that when $\varepsilon=0$, one obtains  $\tilde{p}=\tilde{p}_t$ at the center however, as we reach the surface of the stellar space we can see that   $\tilde{p}_t\geq \tilde{p}$.

\begin{figure}
\centering
\subfigure[~The anisotropy $\Delta$  of solution (\ref{sol})]{\label{fig:Ani}\includegraphics[scale=.3]{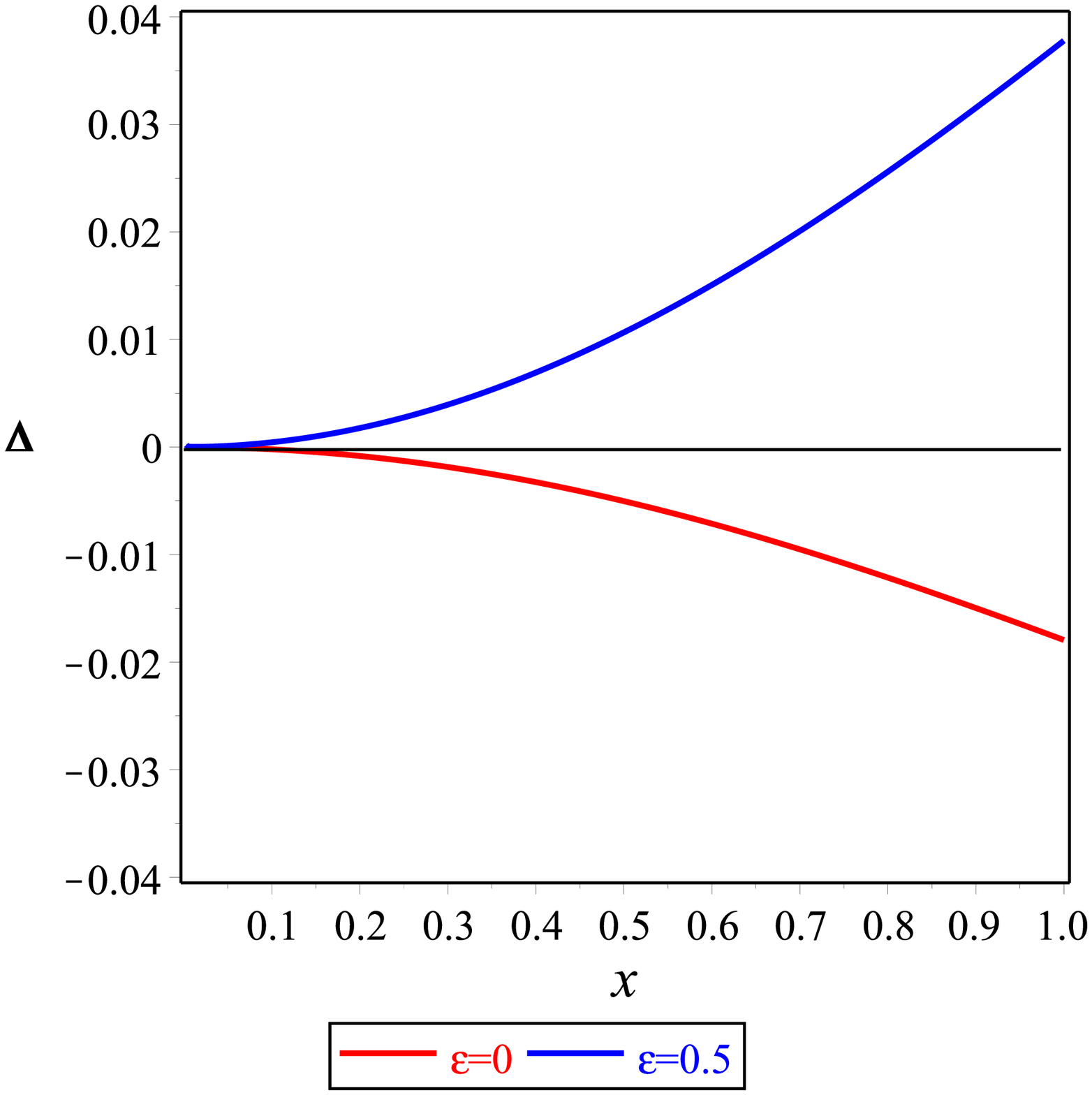}}
\subfigure[~The anisotropic force $\Delta/\chi$  of solution (\ref{sol})]{\label{fig:AnF}\includegraphics[scale=.3]{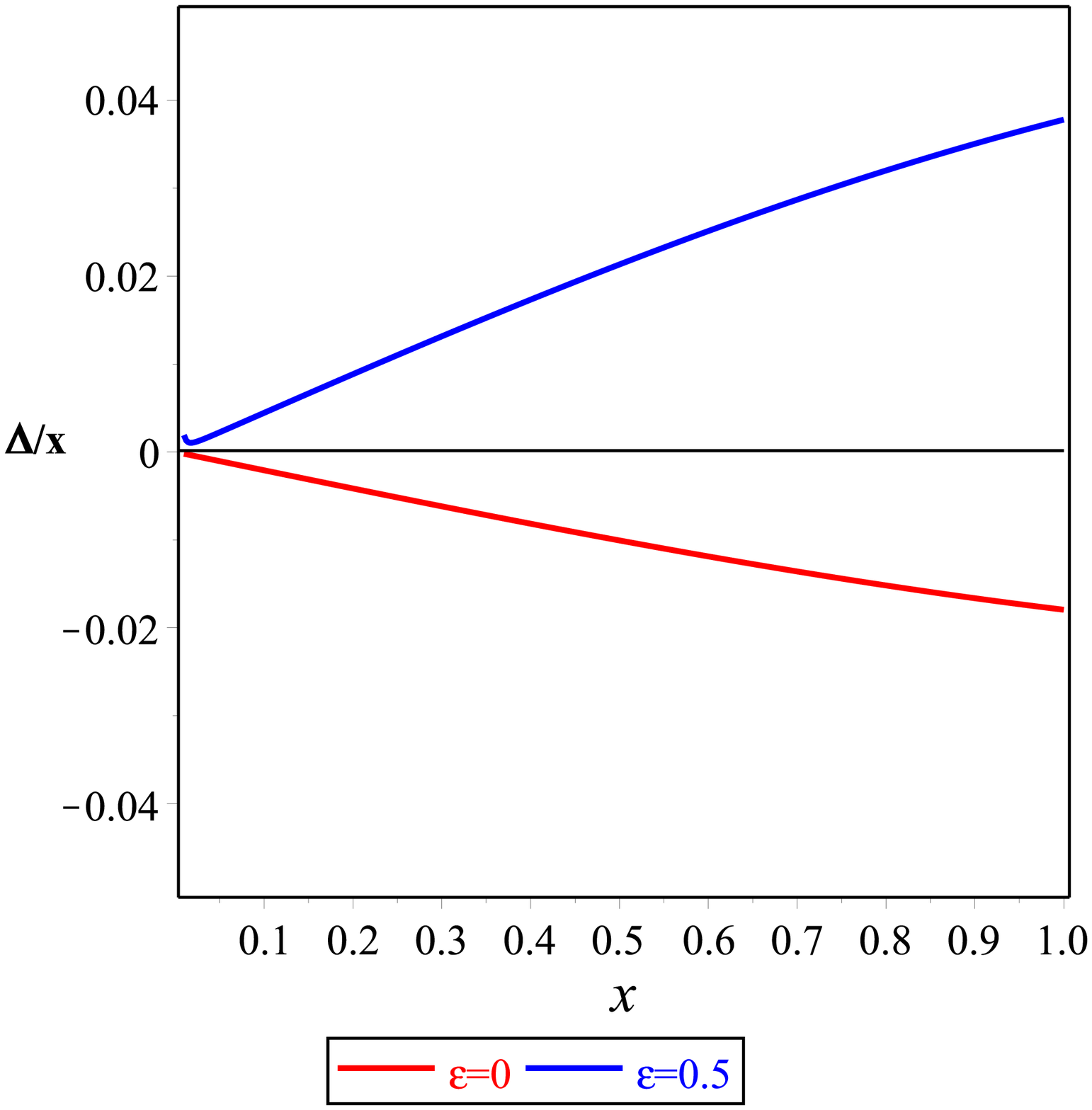}}
\caption[figtopcap]{\small{{Anisotropy $\Delta(\chi)$  and anisotropy force when $\varepsilon=0$ and $\varepsilon=0.5$.}}}
\label{Fig:2}
\end{figure}

  In  Fig. \ref{Fig:2}, we  plot the behavior of the anisotropy and anisotropic force which are defined as $\Delta(\chi)=p_t-p_r$, $F=\Delta(\chi)/\chi$, respectively, for different values of $\varepsilon$. As Fig.  \ref{Fig:2} \subref{fig:Ani} shows  the anisotropic  is positive when $\varepsilon=0.5$ which means that it possesses a repulsive, gravitational, outward force, due to the fact that $p_t\geq p_r$ which means that it enables the formation
of supermassive stars. When  $\varepsilon=0$ ,the anisotropic force becomes negative,  which means that it possesses an inward  gravitational force due to the fact that $p_r\geq p_t$.


\begin{figure}
\centering
\subfigure[~The gradient of energy-density, radial and tangential pressures  of solution (\ref{sol}) when $\varepsilon=0$]{\label{fig:gdnesity}\includegraphics[scale=0.3]{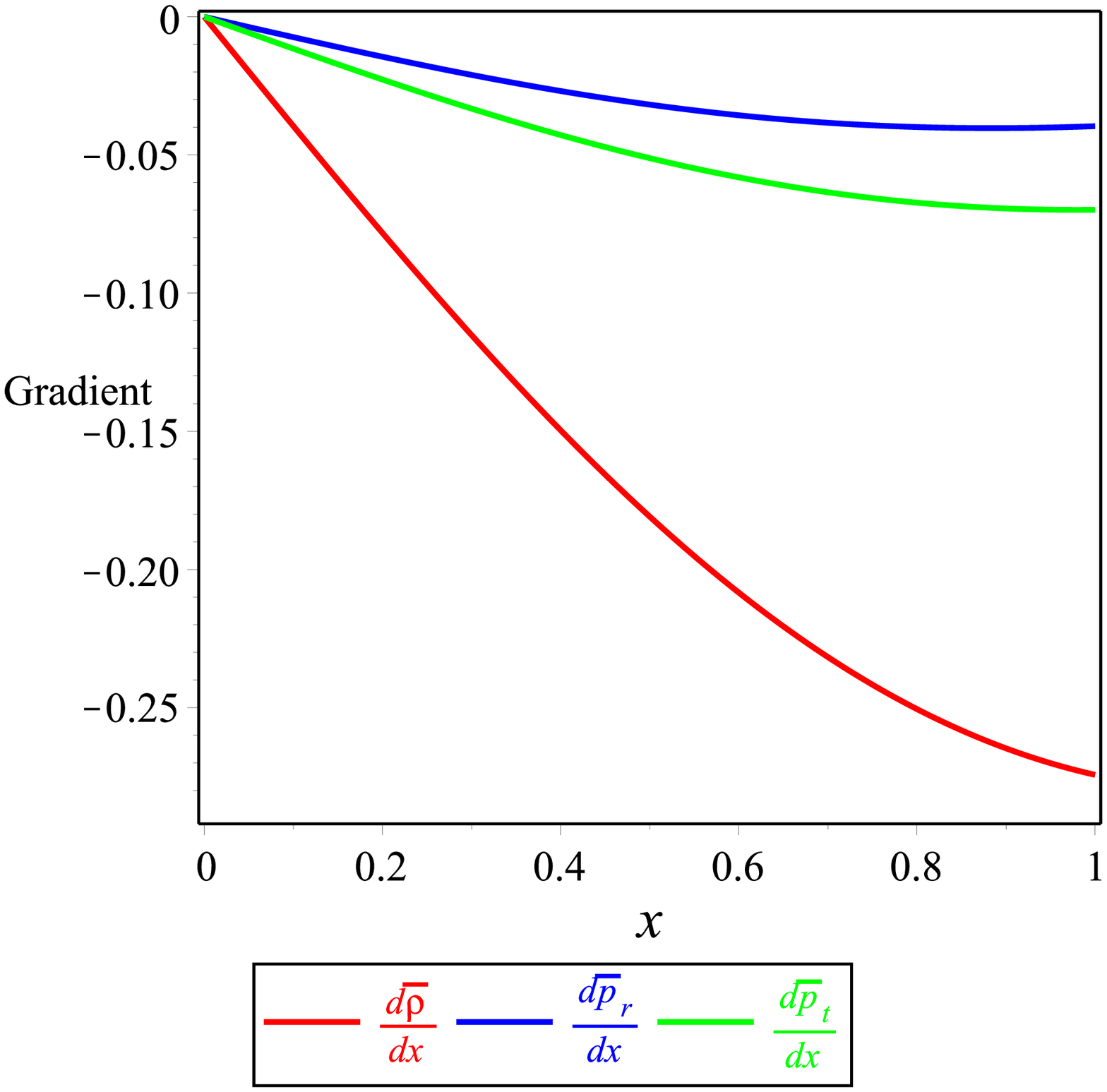}}
\subfigure[~The gradient of energy-density, radial and tangential pressures  of solution (\ref{sol}) when $\varepsilon=0.5$]{\label{fig:gpressure}\includegraphics[scale=0.3]{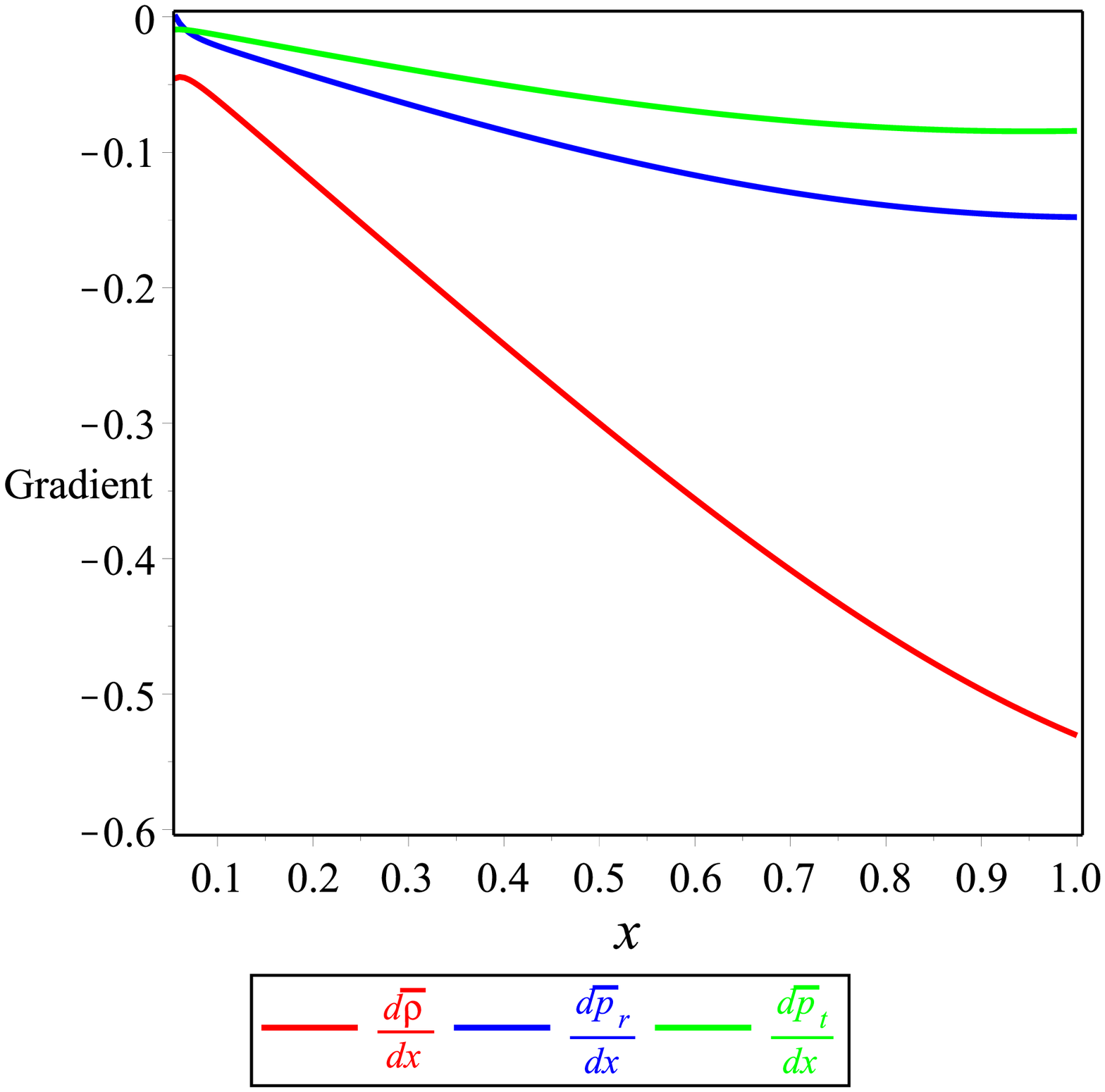}}
%
\caption[figtopcap]{\small{{Gradients of the density, radial, and transverse pressures  when $\varepsilon=0$ and $\varepsilon=0.5$}}}
\label{Fig:3}
\end{figure}
\subsection{\bf Causality}
To show the behavior of the sound velocities we must calculate the gradient of the  energy--density, radial and transverse  pressures which  take the following form:
\begin{eqnarray} \label{grad}
 && \tilde{\rho}'=-\frac{2e^{-2f_2\chi^2}}{\chi^5}
 \Bigg[4\varepsilon e^{f_2\chi^2/2}\Big\{2-3f_0f_2 \chi^6(2f_0-5f_2)-12f_2\chi^4(f_0-f_2)-\chi^2(8f_0-11f_2)\Big\}+e^{3f_2\chi^2/2}\Big(2\varepsilon[4+f_2\chi^2]\nonumber\\
 &&+e^{f_2\chi^2/2}[\chi^2-2\varepsilon]\Big)+e^{f_2\chi^2}\Big\{\varepsilon(8f_2\chi^4[f_0-f_2]-
 4[3-f_0f_2(f_0-f_2)\chi^6]+2\chi^2[4f_0-7f_2])-f_2\chi^4(1-2f_2\chi^2)-\chi^2\Big\}\nonumber\\
 &&+2\varepsilon\Big\{4f_0f_2(f_0-4f_2)\chi^6+8f_2\chi^4(f_0-f_2)+\chi^2(4f_0-5f_2)-1\Big\}\Big]\,,\nonumber\\
 &&\tilde{p}'_r=-\frac{2e^{-2f_2\chi^2}}{\chi^5}
 \Bigg[4\varepsilon e^{f_2\chi^2/2}\Big\{4+6f_0{}^2f_2 \chi^6+9f_0f_2\chi^4+3(2f_0+f_2)\chi^2\Big\}-4e^{2f_2\chi^2}(\chi^2-2\varepsilon)-e^{f_2\chi^2}\Big[12\varepsilon+\chi^2\Big\{2f_0f_2\chi^4\Big(2f_0\varepsilon\nonumber\\
 &&-1\Big)+\chi^4(12f_0f_2\varepsilon-f_2)
 +6\varepsilon(2f_0+f_2)\Big\}\Big]-6\varepsilon(1+2f_0\chi^2)(1+f_2\chi^2[1+2f_0\chi^2])\Big]\,,\nonumber\\
 && \tilde{p}'_t=-\frac{2e^{-2f_2\chi^2}}{\chi^5}
 \Bigg[\varepsilon e^{f_2\chi^2/2}\Big\{3f_0{}^2f_2 \chi^8(2f_0-3f_2)+4f_0\chi^6(9f_0f_2-6f_2{}^2-2f_0{}^2)-12f_2\chi^4(f_2-2f_0)+\chi^2(16f_0-11f_2)-4\Big\}\nonumber\\
 &&+\varepsilon e^{3f_2\chi^2/2}(4+2f_0f_2\chi^4+\chi^2[4f_0+f_2])-2\varepsilon  e^{2f_2\chi^2}+\chi^2 e^{f_2\chi^2}[f_0f_2\chi^6(f_0-f_2)+\chi^4\{2f_0f_2\varepsilon(2f_2-3f_0)+f_2(3f_0-f_2)-f_0{}^2\}\nonumber\\
 &&-2\varepsilon\,(f_2\chi^2+1)(7f_0-2f_2)]+2\varepsilon\{1-
 4f_0{}^2f_2\chi^8(f_0-2f_2)+2f_0\chi^6(f_0{}^2+6f_2{}^2-9f_0f_2)+2f_2\chi^4(3f_0-f_2)+3\chi^2(f_0-f_2)\}\Bigg]\,,\nonumber\\
 &&
  \end{eqnarray}
  where $\tilde{\rho}'=\frac{d\tilde{\rho}}{d\chi}$,  $\tilde{p}'_r=\frac{d\tilde{p_r}}{d\chi}$ and $\tilde{p}'_t=\frac{d\tilde{p}_t}{dx}$. Before graphically analyzing the gradient of the energy-momentum components given by Eq. (\ref{grad})  we will write the asymptote form of Eq. (\ref{grad}) which  yields the following:
  \begin{eqnarray} \label{grad1}
 && \tilde{\rho}'\approx -f_2{}^2\chi(4\varepsilon[f_2-4f_0]+5)+\cdots\,,\qquad \tilde{p}'\approx -f_2\chi(\varepsilon[6f_0f_2-8f_0{}^2-f_2{}^2]+4f_0-f_2)+\cdots\,,\nonumber\\
 &&\tilde{p}'_t\approx  -\frac{\chi(3f_2\varepsilon[6f_0f_2-8f_0{}^2-f_2{}^2]-4f_0{}^2+12f_0f_2-4f_2{}^2)}{2}+\cdots\,,\,,
 \end{eqnarray}
 which gives a negative gradient of the energy-momentum components provided that we use the values of $f_0$ and $f_2$  given by Eq. (\ref{Eq3}). The behavior of the gradient of density, radial and tangential pressures are shown  in Fig. \ref{Fig:3} \subref{fig:gdnesity} and \subref{fig:gpressure} for $\varepsilon=0$ and $\varepsilon=0.5$.

\begin{figure}
\centering
\subfigure[~The radial and transverse speeds  of solution  (\ref{sol}) when $\varepsilon=0$]{\label{fig:dnesity1}\includegraphics[scale=0.3]{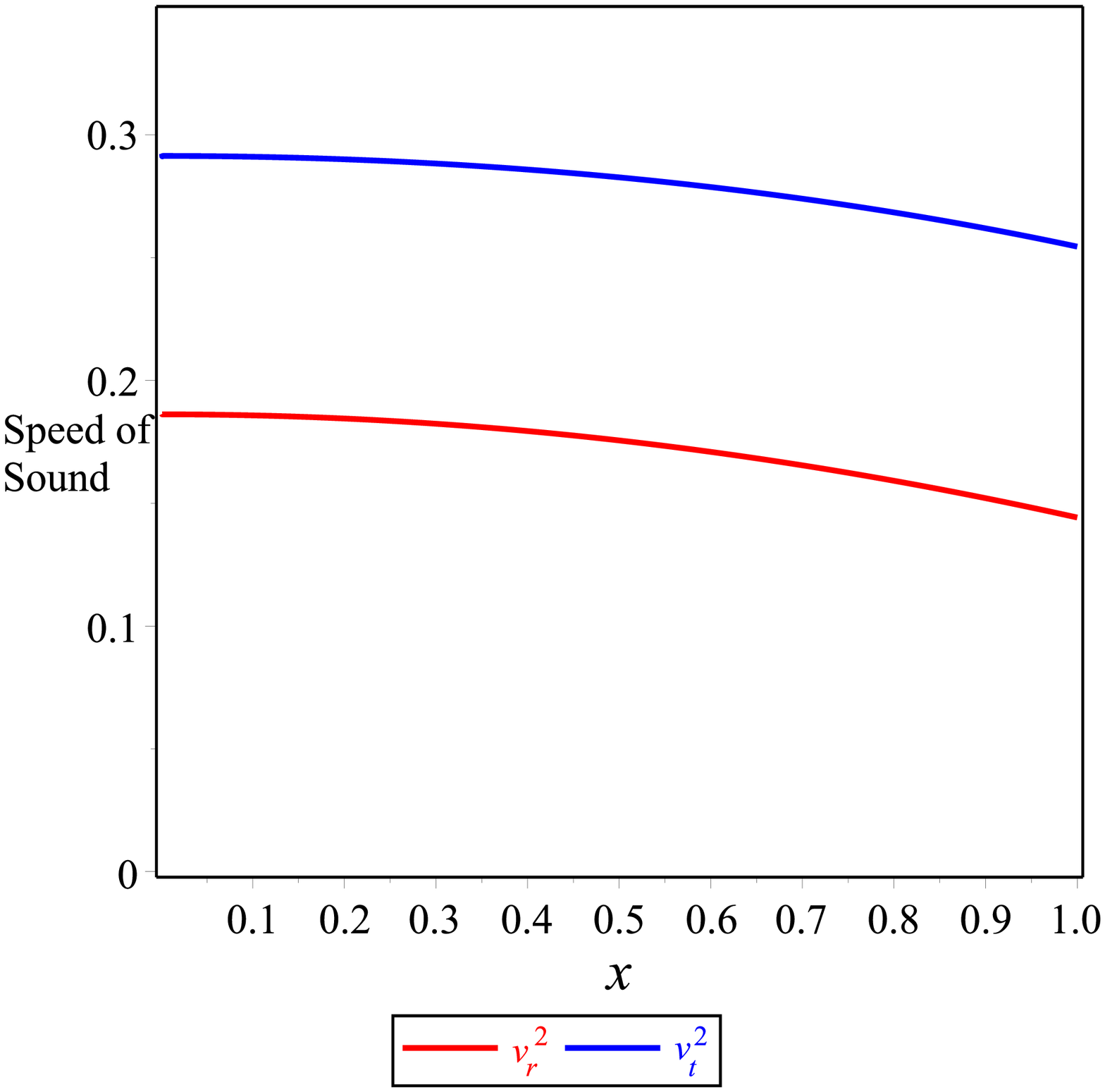}}
\subfigure[~The radial and transverse speeds  of solution  (\ref{sol}) when $\varepsilon=0.5$]{\label{fig:pressure1}\includegraphics[scale=0.3]{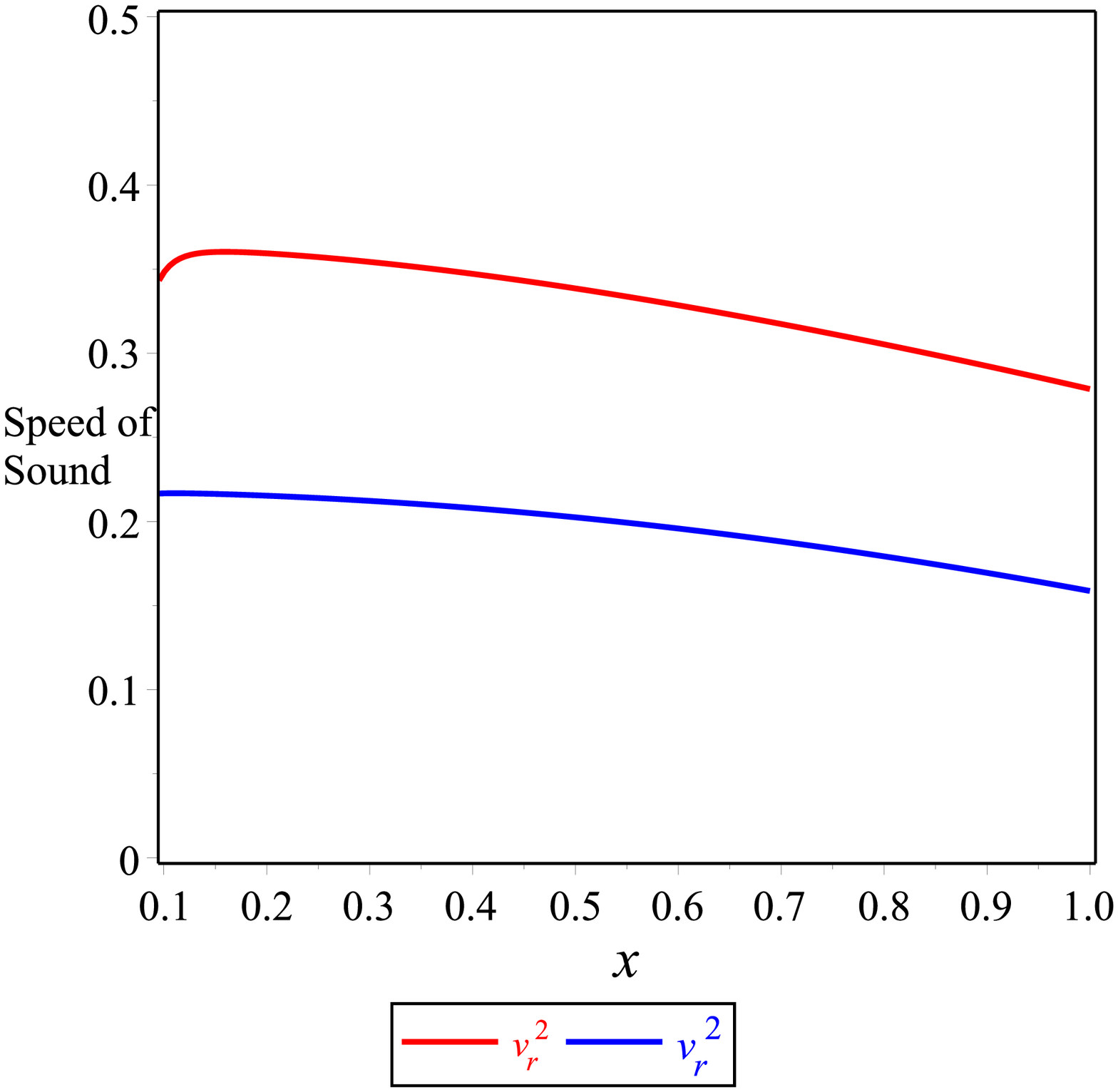}}
\subfigure[~The cracking of  (\ref{sol}) when $\varepsilon=0$ and $\varepsilon=0.5$]{\label{fig:crac}\includegraphics[scale=0.3]{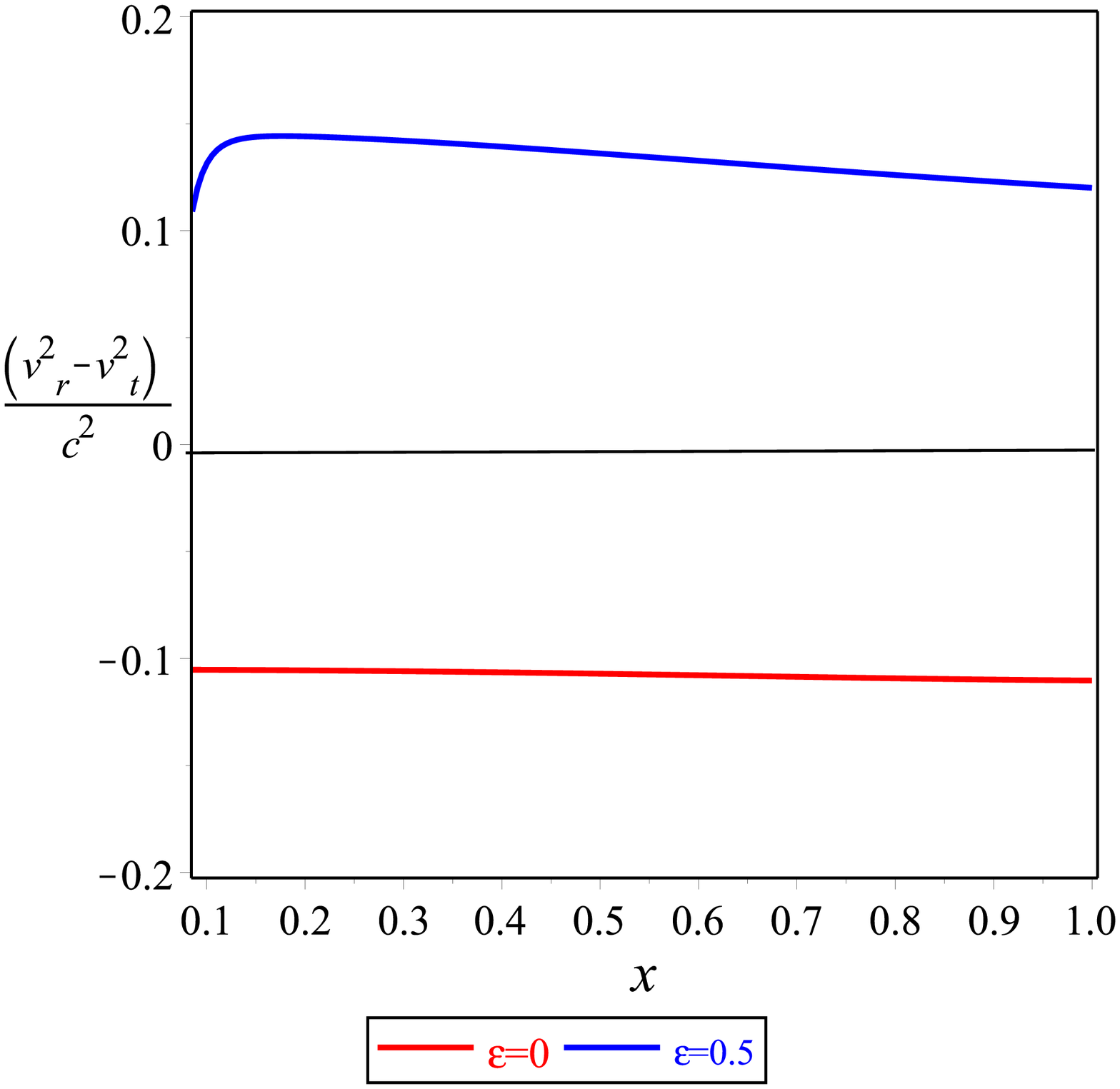}}
\caption[figtopcap]{\small{{Radial, transverse  speeds of sound  when $\varepsilon=0$ and $\varepsilon=0.5$}}}
\label{Fig:4}
\end{figure}

 In order to investigate the causality condition for either the radial or
transverse sound speeds $v_r{}^2$ and $v_\bot{}^2$, we must show that the values of both of them are less than the speed of light.   Using Eq. (\ref{grad}) we obtain the asymptotic form of the radial and transverse speeds as the following:
  \begin{eqnarray} \label{grad2}
 && v_r{}^2=\frac{\tilde{p}'_r}{\tilde{\rho}'}\approx \frac{1}{12f_2(5-4[4f_0-f_2]\varepsilon)^2}\Big[3\varepsilon^2[11f_2{}^3\chi^2+2f_2{}^2(8-19f_0\chi^2)-32f_0f_2(3+f_0\chi^2)-32f_0{}^2(f_0\chi^2-4)](4f_0-f_2)\nonumber\\
 &&-\varepsilon[30f_0{}^3\chi^2+f_2{}^2(108-196f_0\chi^2)-f_2(744f_0-
 592f_0{}^2\chi^2)-96f_0{}^2(2f_0\chi^2-13)]+60(4f_0-f_2)-16(f_0+f_2)f_2\chi^2\Big]\,, \nonumber\\
 &&
 v_t{}^2=\frac{\tilde{p}'_t}{\tilde{\rho}'}\approx\frac{1}{24f_2{}^2(5-4[4f_0-f_2]\varepsilon)^2}\Big[57\varepsilon^2f_2{}^5\chi^2+6f_2{}^4
 \varepsilon[10\chi^2-3\varepsilon(8-47f_0\chi^2)]+4f_2{}^3[4\chi^2-\varepsilon(93-123f_0\chi^2)] \nonumber\\
 &&-f_2{}^2(96\varepsilon^2 f_0{}^2[48-73f_0\chi^2]+12\varepsilon f_0[202+37f_0 \chi^2]-48[6f_0\chi^2+5])+f_2(384\varepsilon^2 f_0{}^3[12+5f_0 \chi^2] \nonumber\\
 &&-96\varepsilon f_0{}^2[41-9f_0 \chi^2]+16f_0[45+16f_0 \chi^2])+6f_0{}^3 \varepsilon [131-32 f_0 \chi^2]-240 f_0{}^2\Big]\,.
 \end{eqnarray}
 Equation (\ref{grad2}) shows that both $c^2>v_r{}^2>0$ and $c^2>v_t{}^2>0$ utilize the data presented in Eq. (\ref{Eq3}). The behavior of the radial and tangential speeds of sound are shown in Fig. \ref{Fig:4} \subref{fig:dnesity1} and \subref{fig:pressure1}.

 The appearance of the non--vanishing total radial force with
different signs in different regions of the
fluid is called
gravitational cracking when this radial force is directed
inward to the inner part of the sphere for all values of the
radial coordinate $\chi$ between the center and some value
beyond which the force reverses its direction \cite{1994PhLA..188..402H}. In
 \cite{Abreu:2007ew} it is stated that a simple requirement for
avoiding gravitational cracking is the following $0< v_r{}^2-v_t{}^2<c^2$. In Fig.  \ref{Fig:4} \subref{fig:crac} we demonstrate  that solution (\ref{sol}) is locally  stable against cracking for  $\varepsilon=0.5$  but not for $\varepsilon=0$.
\subsection{\bf Energy conditions}
\begin{figure}
\centering
\subfigure[~The DEC,  $\tilde{\rho}-\tilde{p}_r$, of solution  (\ref{sol})]{\label{fig:Decr}\includegraphics[scale=0.3]{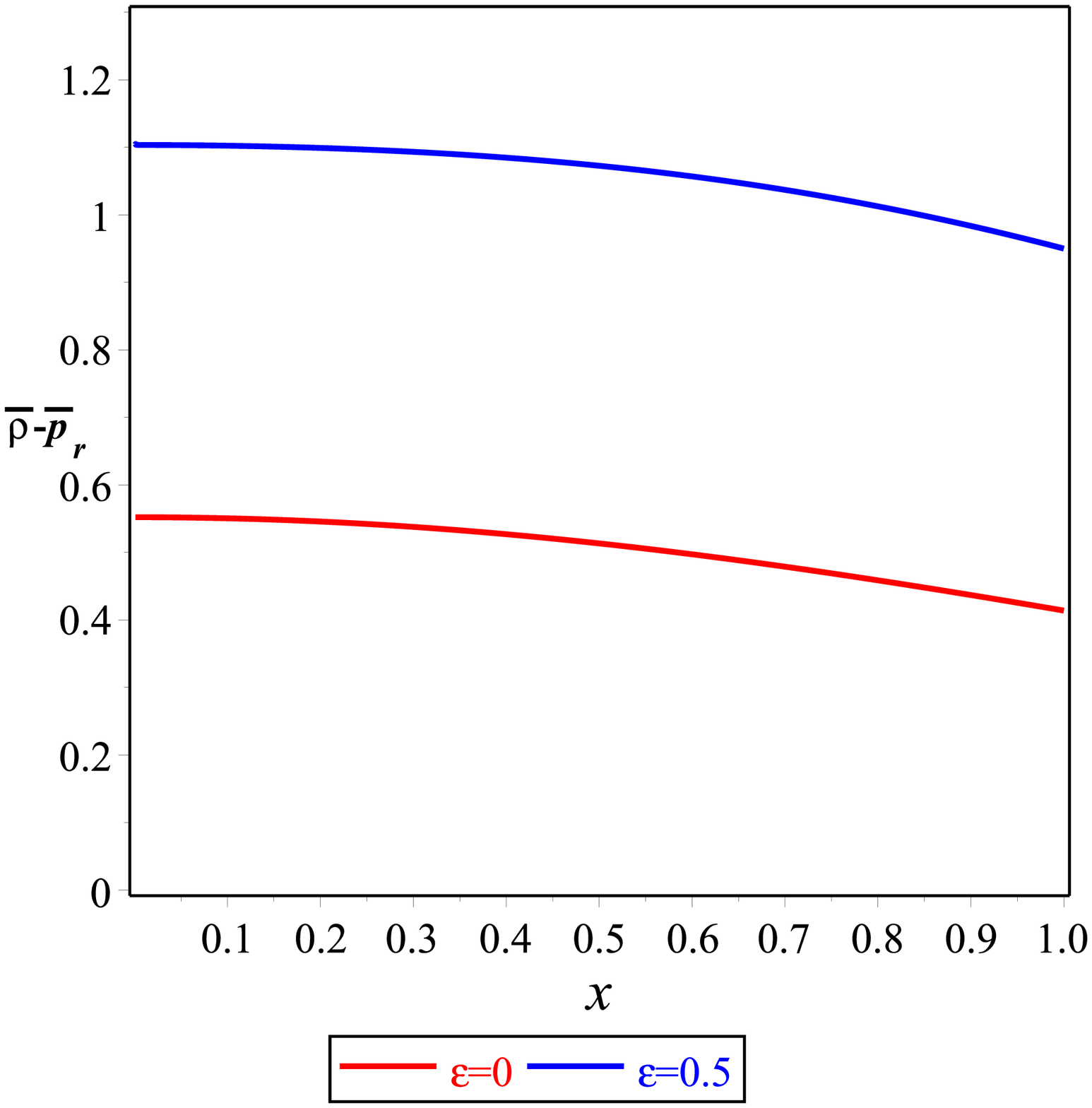}}
\subfigure[~The DEC,   $\tilde{\rho}-\tilde{p}_t$, of solution  (\ref{sol})]{\label{fig:Dect}\includegraphics[scale=.3]{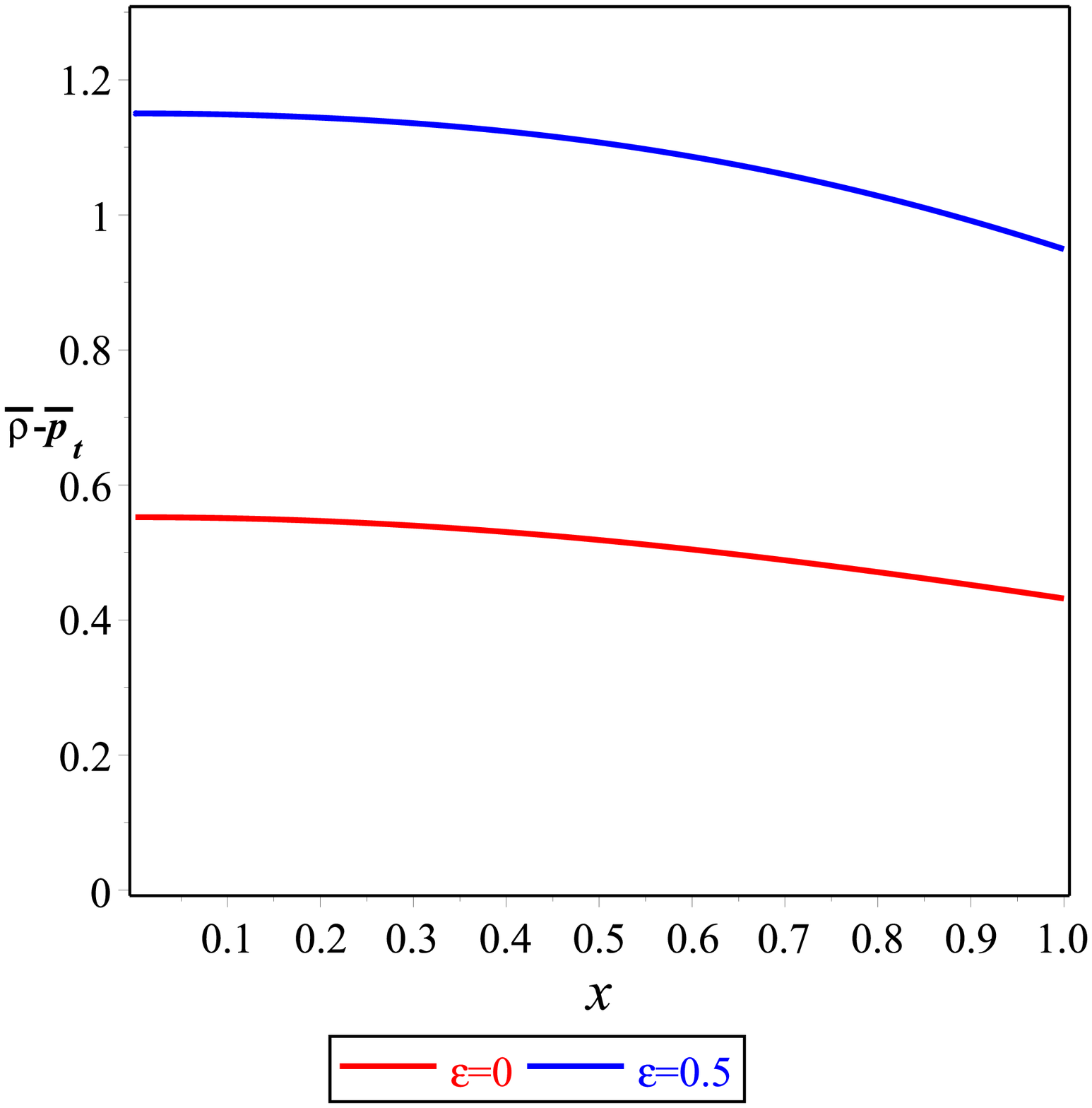}}
\subfigure[~The WEC, $\tilde{\rho}+\tilde{p}_r$ of solution  (\ref{sol})]{\label{fig:WEC}\includegraphics[scale=.3]{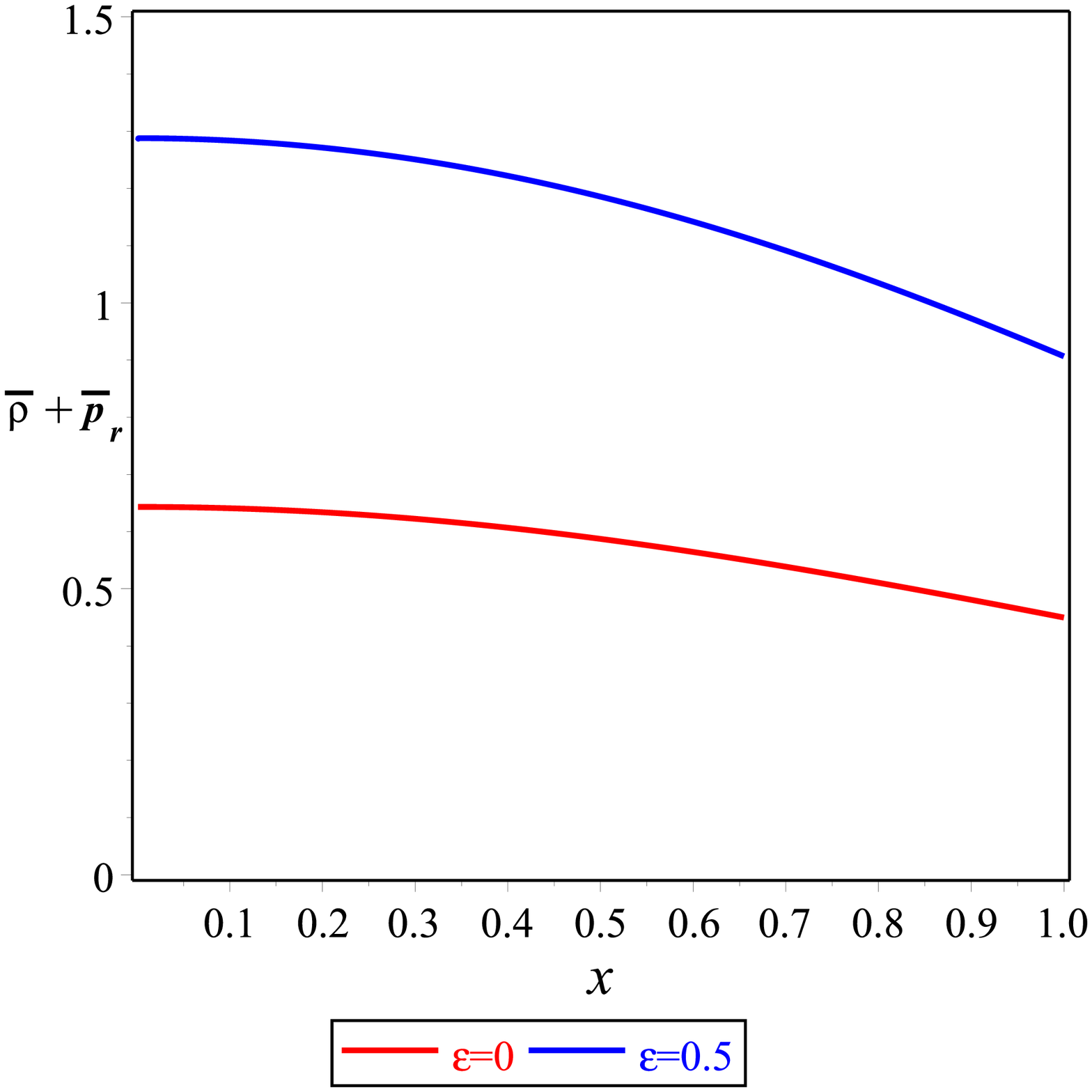}}
\subfigure[~The WEC $\tilde{\rho}+\tilde{p}_t$ of solution  (\ref{sol})]{\label{fig:WEC1}\includegraphics[scale=.3]{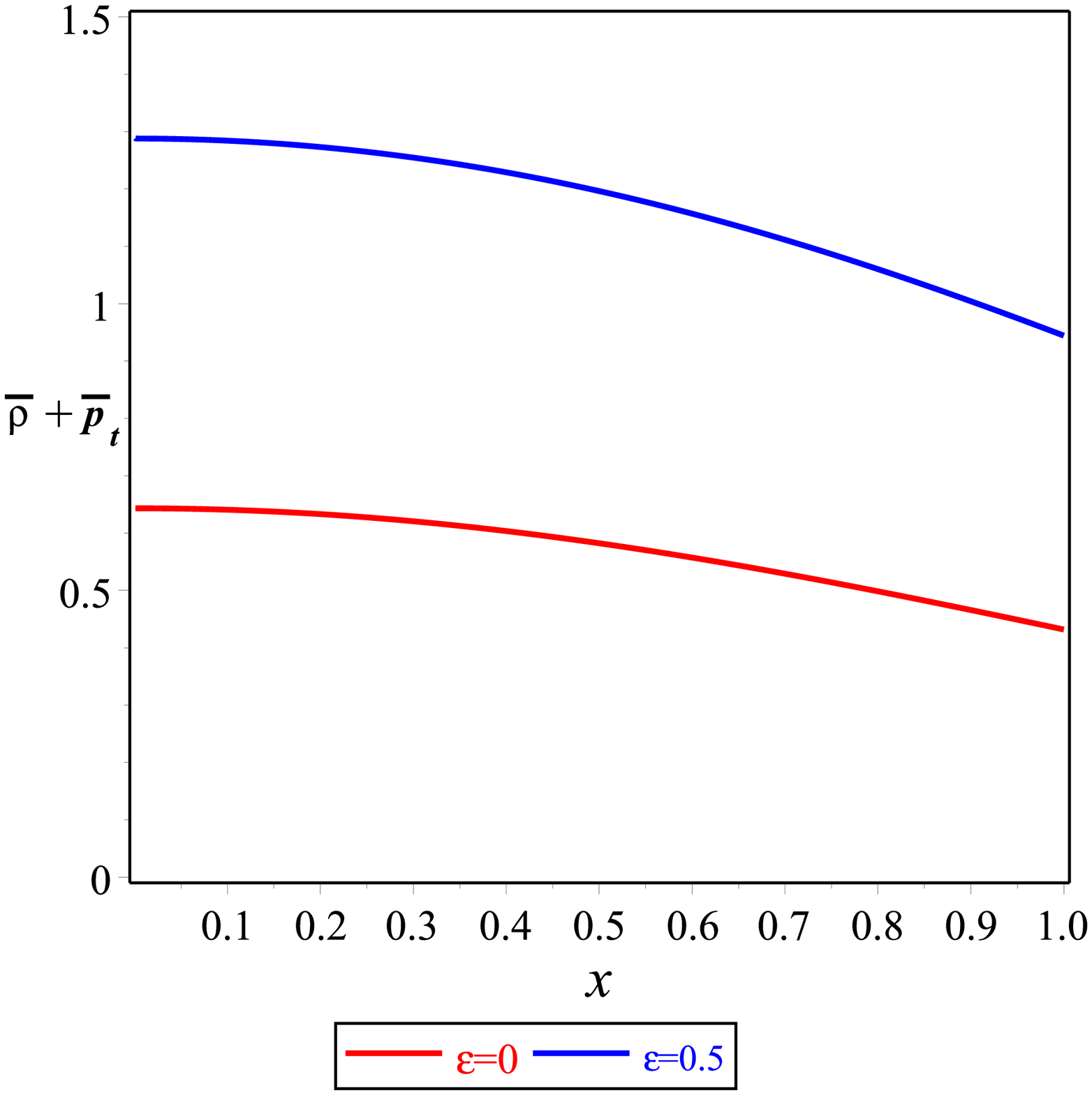}}
\subfigure[~The SEC  $\tilde{\rho}-\tilde{p}_r-2\tilde{p}_t$ of solution  (\ref{sol})]{\label{fig:SEC}\includegraphics[scale=.3]{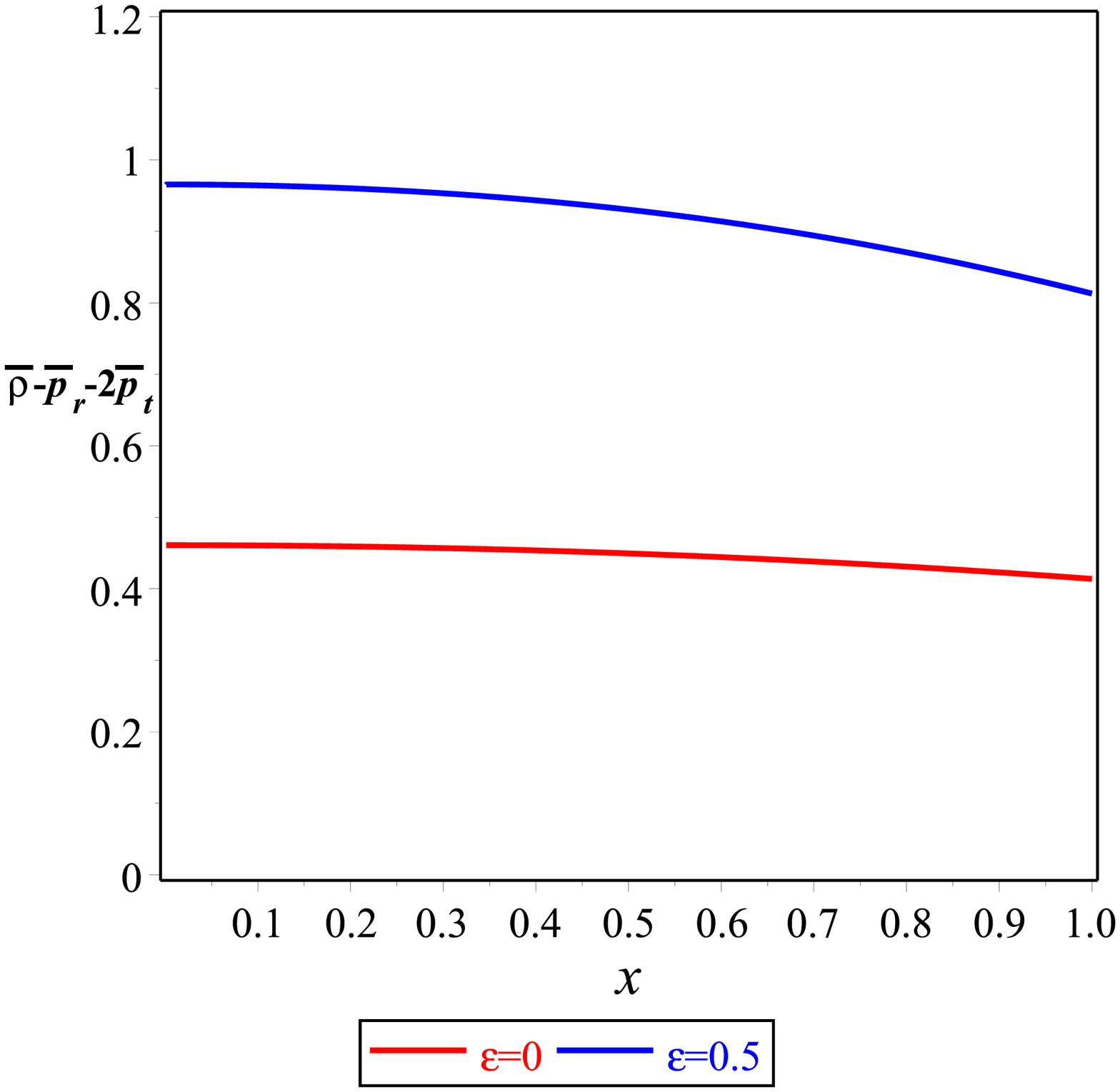}}
\caption[figtopcap]{\small{{ Energy conditions of solution  (\ref{sol})  when $\varepsilon=0$ and $\varepsilon=0.5$}}}
\label{Fig:5}
\end{figure}

The energy conditions are  an important test for the non--vacuum solution. The dominant energy condition (DEC)  leads to the fact that the speed of the energy must be less than that of light.
The asymptotic form of the DEC, strong energy condition  and weak energy condition  take the following forms:
\begin{eqnarray} \label{ECA}
 && DEC\quad \left\{
                \begin{array}{l} \tilde{\rho}-\tilde{p}_r\approx 2(2f_2-f_0)-1/2\chi^2(2f_2{}^2+4f_0f_2-4f_2[2f_0-f_2]+\varepsilon f_2[5f_2{}^2-22f_0f_2+8f_0f_2])+\cdots\,,\\
                 \\
                 \tilde{\rho}-\tilde{p}_t\approx 2(2f_2-f_0)-1/2\chi^2(3f_2{}^2+2f_0(f_0+f_2)-4f_2[2f_0-f_2]+\varepsilon
                 f_2[11f_2{}^2-25f_0f_2+12f_0f_2])+\cdots\,,\nonumber\\
                  \end{array}
              \right.
              \\
              \\
 &&  WEC \quad \left\{
                \begin{array}{l} \tilde{\rho}+\tilde{p}_r\approx 2(f_2+f_0)-1/2\chi^2(8f_2{}^2-4f_0f_2+4f_2[2f_0-f_2]+\varepsilon f_2[3f_2{}^2-10f_0f_2-8f_0f_2])+\cdots\,,\\
                 \\
                 \tilde{\rho}-\tilde{p}_t\approx 2(f_2+f_0)-1/4\chi^2(14f_2{}^2-4f_0(f_0+f_2)+8f_2[2f_0+f_2]+\varepsilon f_2[5f_2{}^2-14f_0f_2-24f_0f_2])+\cdots\,,\nonumber\\
  \end{array}
              \right.
              \\
              \\
&&  SEC \quad \tilde{\rho}-\tilde{p}_r-2\tilde{p}_t\approx 6(f_2-f_0)+\chi^2(f_2{}^2+6(2f_0-f_2)f_2-2f_0[2f_2-f_0]+4\varepsilon f_2[5f_0f_2-f_2{}^2-4f_0f_2])+\cdots\,.\nonumber\\
  &&
\end{eqnarray}
All of the above energy conditions are analytically and  graphically satisfied using Eq. (\ref{Eq3}). Figure  \ref{Fig:5} shows the behavior of the energy conditions  for vanishing and non-vanishing values of the dimensional parameter  $\rho$.
\subsection{\bf Mass--radius--relation}
\begin{figure}
\centering
\subfigure[~The gravitational mass of  (\ref{sol}) when $\varepsilon=0$ ]{\label{fig:mass}\includegraphics[scale=0.3]{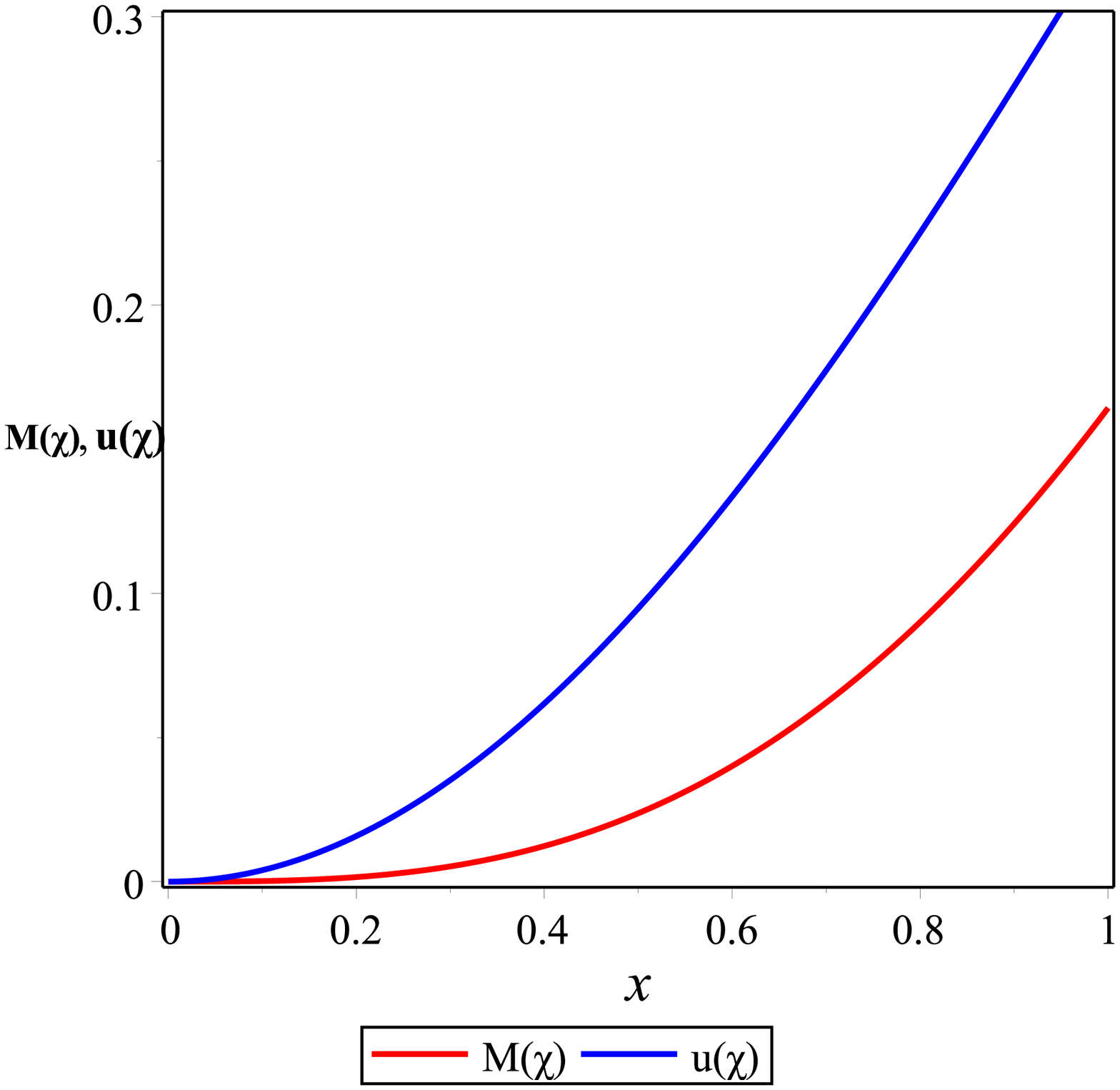}}
\subfigure[~The compactification factor of  (\ref{sol}) when  $\varepsilon=0.5$]{\label{fig:u}\includegraphics[scale=.3]{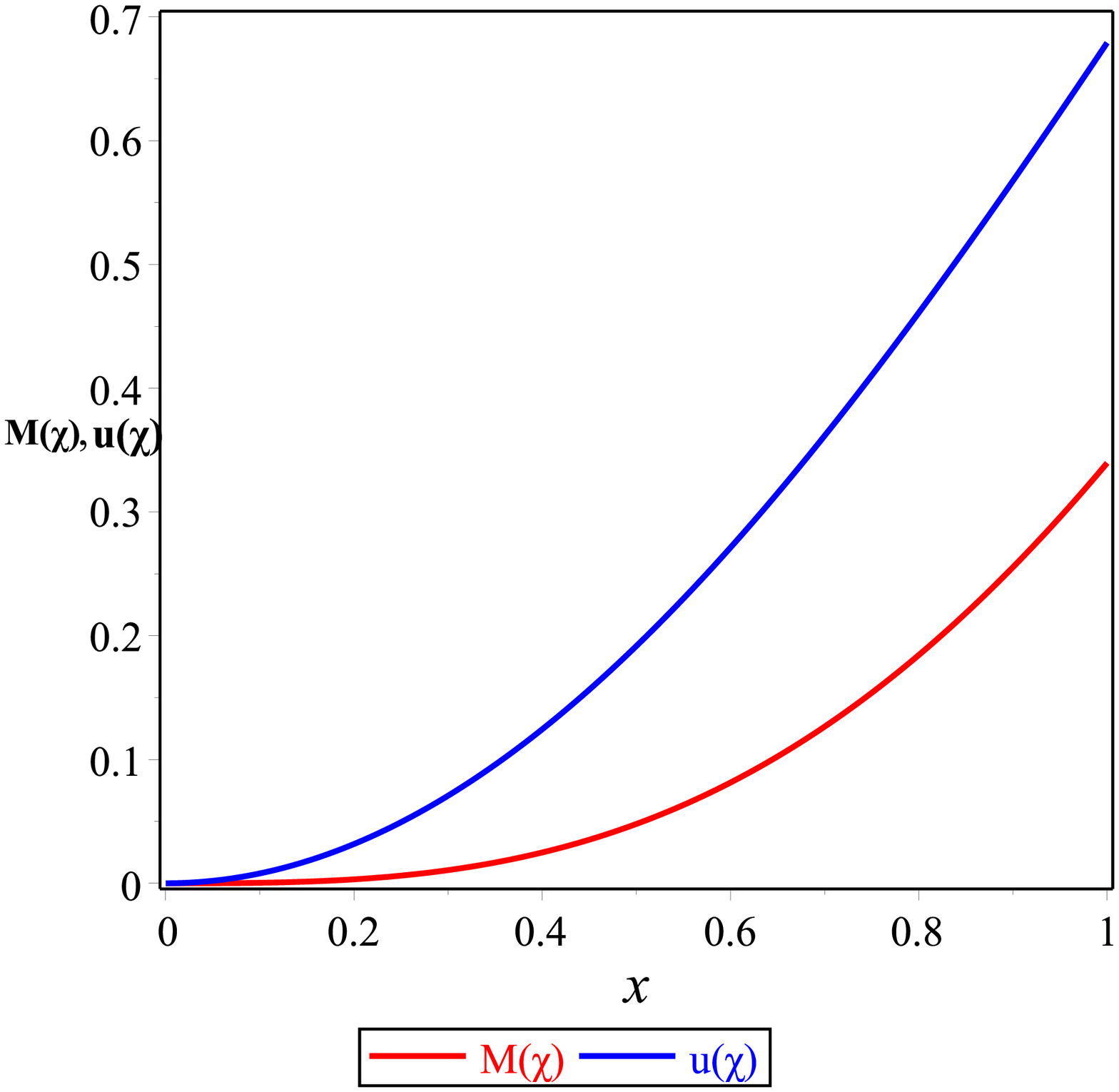}}
%
\caption[figtopcap]{\small{{Gravitational mass and compactification factor  of solution  (\ref{sol}) when $\varepsilon=0$ and $\varepsilon=0.5$}}}
\label{Fig:6}
\end{figure}

The compactification factor $u(r)$ of a compact stellar
 is defined as the ratio between the mass and radius and is considered as  an ingredient role to use  to understand the physical properties
of the star. From solution (\ref{sol}) we derived  the gravitational mass as the following:
 \begin{eqnarray}\label{mass}
&& M(\chi)={\int_0}^\chi  \tilde{\rho(\xi)} \xi^2 d\xi\nonumber\\
 &&=\frac{1}{36f_0{}^{3/2}\chi}\Bigg[8\chi\varepsilon \sqrt{6\pi}\,erf(\sqrt{6f_2}\,\chi/2)(3f_2{}^2-2f_2{}^2-7f_0f_2)+48\varepsilon e^{-3/2f_2\chi^2}\, \sqrt{f_2}(2f_0{}^2\chi^2-f_2[3+5f_0\chi^2])-144\varepsilon f_2{}^{3/2}e^{-f_2 \chi^2/2}\nonumber\\
 &&-72\chi\varepsilon f_2{}^2\sqrt{2\pi}\,erf(\sqrt{2f_2}\,\chi/2)-
36 e^{-f_2 \chi^2}\sqrt{f_2}[2\varepsilon f_0{}^2 \chi^2+f_2(\chi^2-2\varepsilon\{3-f_0\chi^2\})-\varepsilon e^{-f_2 \chi^2}(f_0{}^2 \chi^2+f_2\{1-4\chi^2f_0\})]\nonumber\\
 &&+\varepsilon \chi \sqrt{2\pi}\,erf(\sqrt{2f_2}\,\chi)[f_0{}^2+4f_0f_2-4f_2{}^2]+36\varepsilon \chi \sqrt{\pi}(f_0+2f_2)(f_0+f_2)\,erf(\sqrt{f_2}\,\chi)+36f_2{}^{3/2}(\chi^2+\varepsilon)\Bigg]\,,\nonumber\\
 && \approx \frac{f_2 \chi^3}{10}(5[2-f_2\chi^2]-4\varepsilon f_2 \chi^2[f_2-4f_0])\,,
\end{eqnarray}
where $erf(\chi)$ is the error function which  is defined as the following:
\begin{eqnarray}
erf(\chi)=\frac{2}{\sqrt{\pi}}\int_0^x e^{-y^2}dy\,.\end{eqnarray} The compactification factor $u(\chi)$  which is defined as the following:
\begin{eqnarray}\label{comp}
&&u(\chi)=\frac{M(\chi)}{R}\approx \frac{f_2 \chi^3}{10R}(5[2-f_2\chi^2]-4\varepsilon f_2 \chi^2[f_2-4f_0])\,,\end{eqnarray} where $R$ is the radius of the star and we  use Eq. (\ref{mass}) in (\ref{comp}).
The behavior of the gravitational mass and the  compactification factor are plotted in Fig. \ref{Fig:6} for $\varepsilon=0$ and $\varepsilon =0.5$.
As Fig. \ref{Fig:6} \subref{fig:mass} shows  the gravitational mass increases as the radial coordinate does whereas Fig. \ref{Fig:6} \subref{fig:u} displays the  compactification factor  in direct proportion to the  radial coordinate.
\subsection{\bf Equation of state}
\begin{figure}
\centering
\subfigure[~Radial and tangential EoS when $\varepsilon=0$]{\label{fig:pr}\includegraphics[scale=0.3]{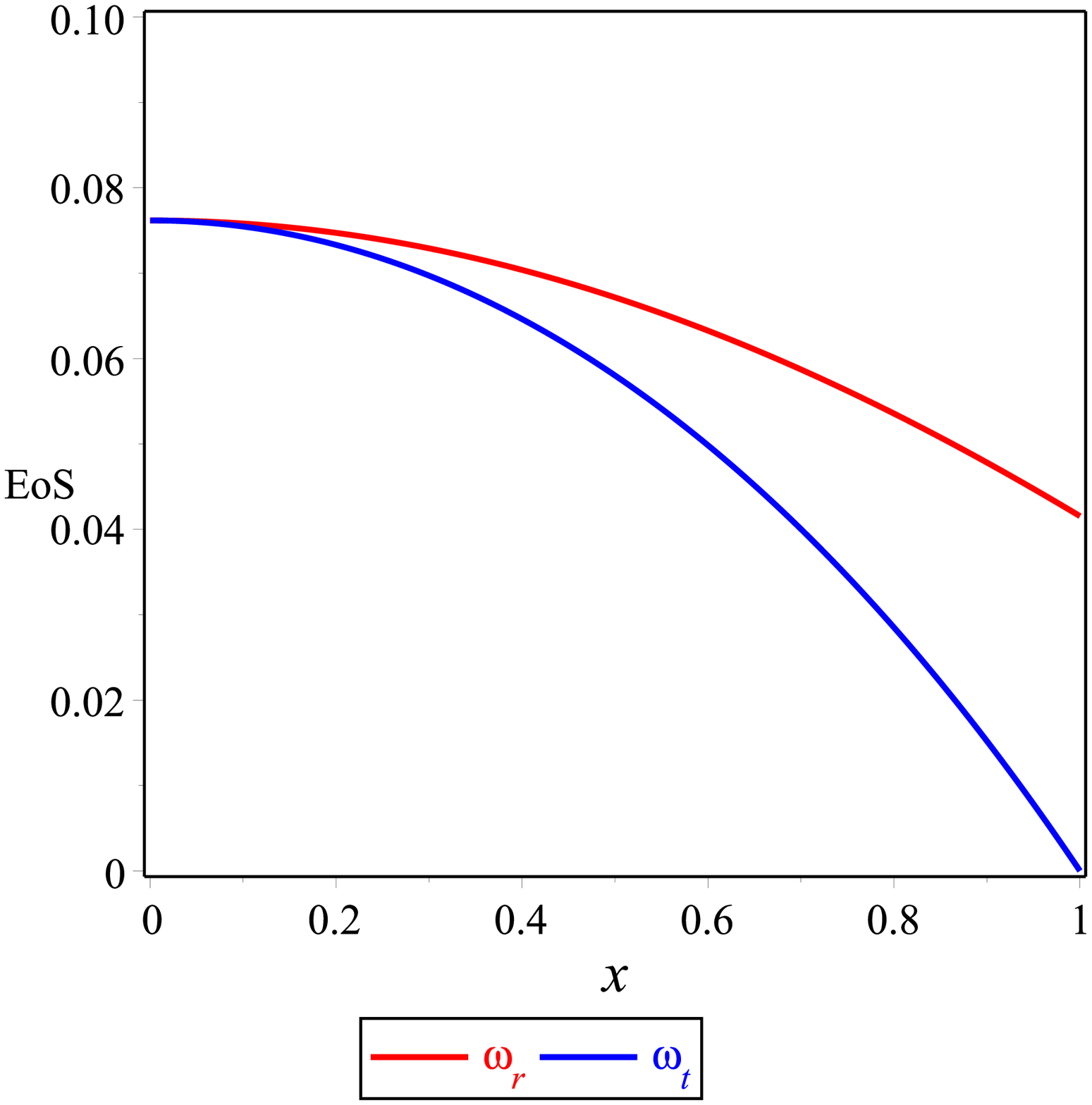}}
\subfigure[~Radial and tangential EoS when $\varepsilon\neq0$]{\label{fig:pt}\includegraphics[scale=.3]{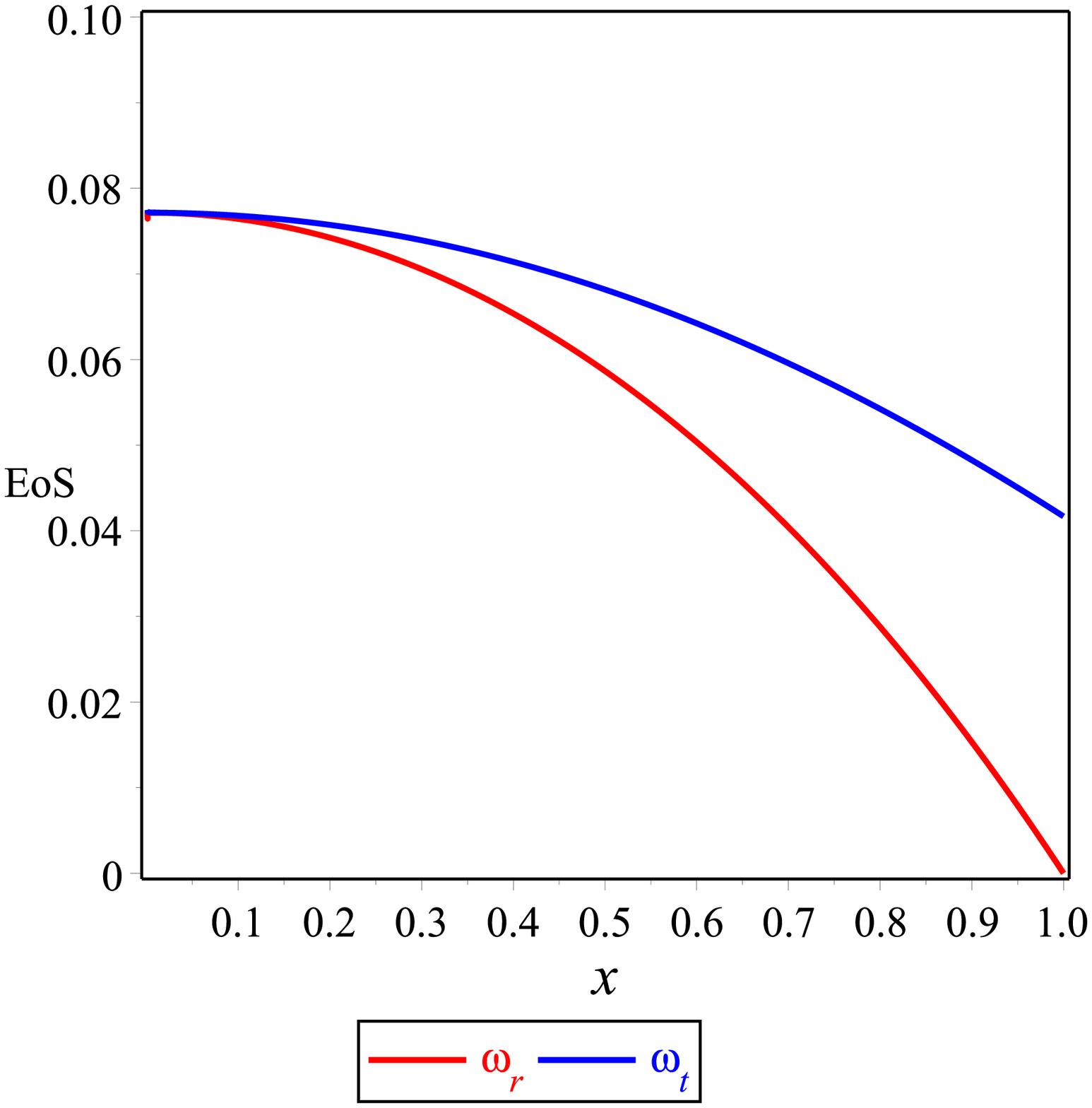}}
\caption[figtopcap]{\small{{Radial and transverse EoS's  of solution  (\ref{sol}) for $\varepsilon=0$ and $\varepsilon=-0.01$}}}
\label{Fig:7}
\end{figure}

Das et al. \cite{Das:2019dkn}  showed that the EoS of neutral compact stars in the frame of GR behaves in a linear form however,  this study  shows that the EoS of solution (\ref{sol}) is not a linear one either for the TEGR or higher order torsion tensor. To rectify  this, we define  the radial and transverse   EoS  as the following:
  \begin{eqnarray} \label{sol3}
 && \omega_r=\frac{\tilde{p}_r}{\tilde{\rho}}\approx \frac{2f_0-f_2}{3f_2}-\frac{2f_2[f_0+f_2]-\varepsilon f_2[6f_0f_2-f_2{}^2-8f_0{}^2]}{18f_2}\chi^2+\cdots\,, \nonumber\\
 && \omega_t=\frac{\tilde{p}_t}{\tilde{\rho}}\approx \frac{4f_0-3f_2}{6f_2}+\frac{2[f_2{}^2-8f_0f_2+6f_0{}^2]+\varepsilon f_2[f_2{}^2-6f_0f_2++8f_0{}^2]}{36f_2}\chi^2+\cdots\,,
  \end{eqnarray}
   where $\omega_r$ and $\omega_t$ are  the radial and transverse.  The behavior of the EoS  is shown in Fig  \ref{Fig:7}.
 As Fig \ref{Fig:7} \subref{fig:pr} and \subref{fig:pt} show that the EoS are also  not linear for the case of TEGR, $\varepsilon=0$ or for the  higher order torsion.  Figure \ref{Fig:7} \subref{fig:pr} and \subref{fig:pt} indicate that for the TEGR case the radial EoS moves to the surface of the star rapidly in comparison to  the  tangential  EoS however for the higher order torsion scalar case the inverse is occurs.
\section{Stability of the model}\label{S6}
The model's stability is an important aspect that any physical model must consider. For this we are going to use  two different methods, the first being   the TOV equation and the second  being the adiabatic index.
\subsection{Tolman-Oppenheimer-Volkoff equation}
\begin{figure}
\centering
\subfigure[~The TOV Equation when $\varepsilon=0$ ]{\label{fig:TOV0}\includegraphics[scale=0.3]{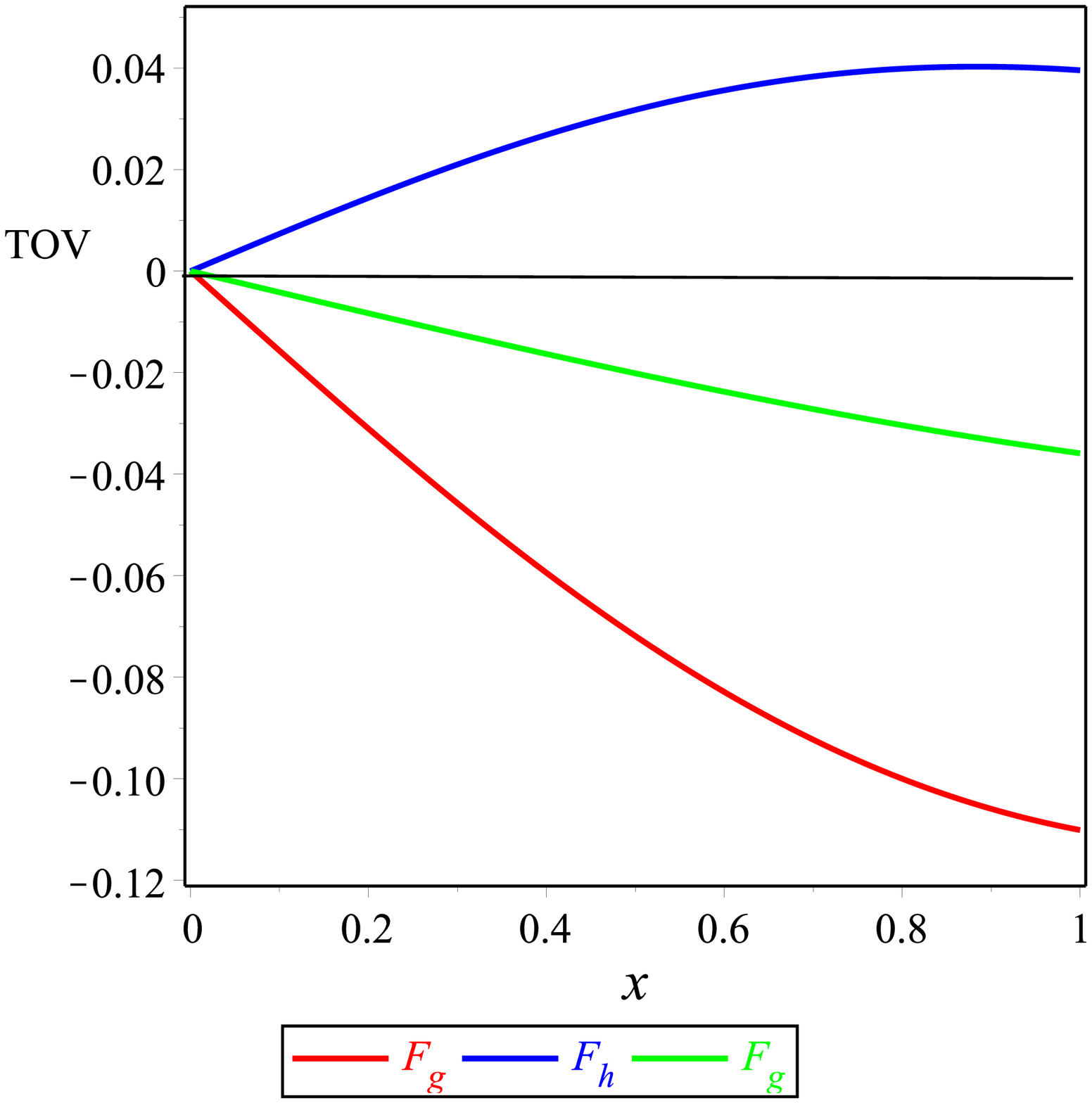}}
\subfigure[~The TOV Equation when $\varepsilon=0.5$]{\label{fig:TOV1}\includegraphics[scale=0.3]{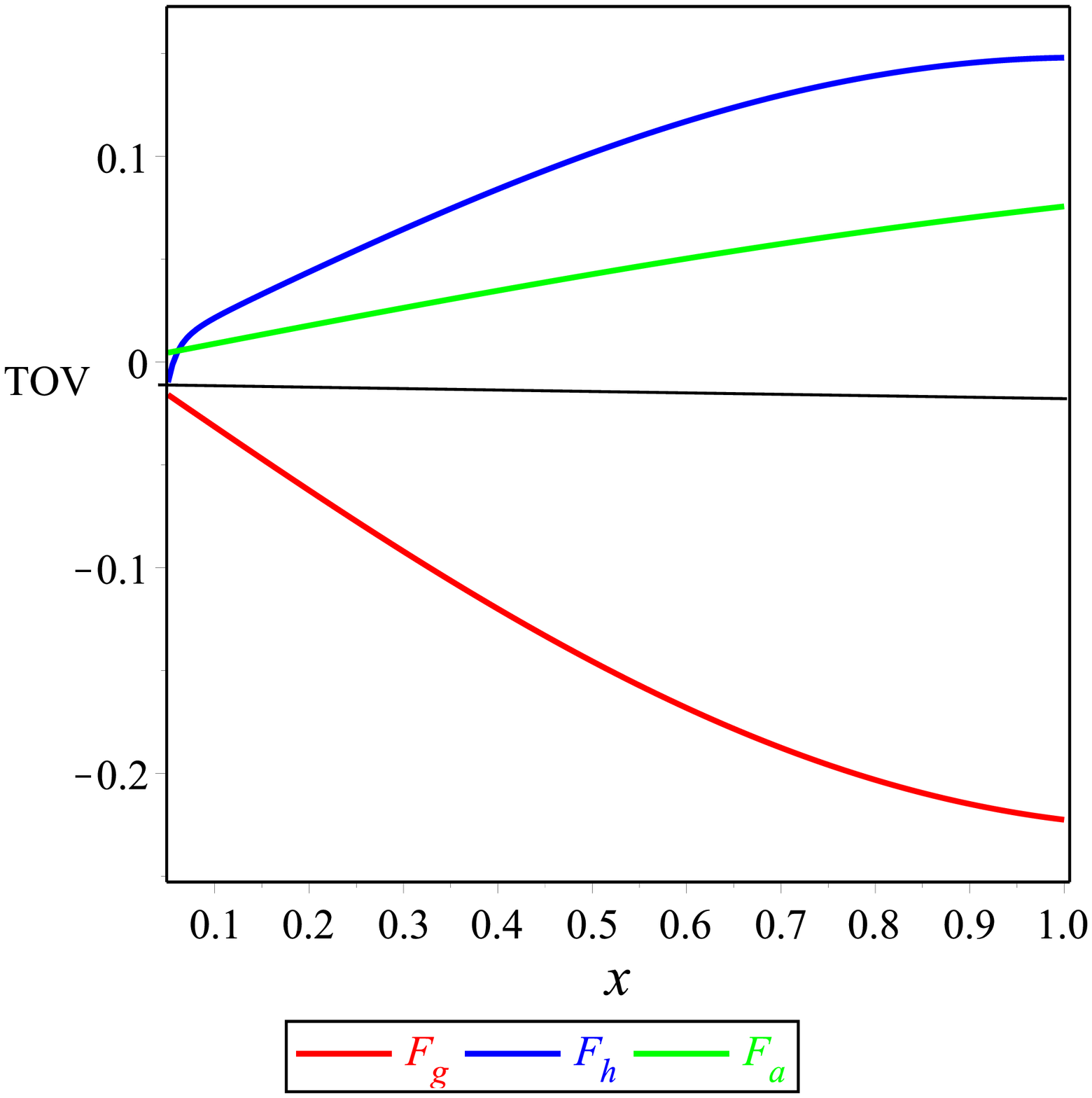}}
\caption[figtopcap]{\small{{TOV of solution  (\ref{sol}) when $\varepsilon=0$ and $\varepsilon=0.5$}}}
\label{Fig:8}
\end{figure}

 We assume a hydrostatic equilibrium in order to study the stability of solution (\ref{sol}). The TOV equation  \cite{PhysRev.55.364,PhysRev.55.374} which  was presented in \cite{PoncedeLeon1993} is as follows:
\begin{eqnarray}\label{TOV} \frac{2[\tilde{p}_t-\tilde{p}_r]}{\chi}-\frac{M_g(\chi)[\tilde{\rho}(\chi)+\tilde{p}_t(\chi)]e^{[\alpha(\chi)-\beta(\chi)]/2}}{\chi}-\frac{d\tilde{p}(\chi)}{\chi}=0.
 \end{eqnarray}
The quantity h $M_g(\chi)$ is the gravitational mass of
radius $\chi$ which is defined as the following:
\begin{eqnarray}\label{ma} M_g(\chi)=4\pi{\int_0}^\chi\Big({{\mathcal T}_t}^t-{{\mathcal T}_r}^r-{{\mathcal T}_\theta}^\theta-{{\mathcal T}_\phi}^\phi\Big)\chi^2e^{[\alpha(\chi)+\beta(\chi)]/2}d\chi=\frac{\chi^2 \alpha' e^{[\beta(\chi)-\alpha(\chi)]/2}}{2}\,,
 \end{eqnarray}
Using Eqs. (\ref{ma}) in (\ref{TOV}) we get
\begin{eqnarray}\label{ma1} \frac{2(\tilde{p}_t-\tilde{p}_r)}{\chi}-\frac{d\tilde{p}_r}{d\chi}-\frac{\alpha'[\tilde{\varepsilon}(\chi)+\tilde{p}_r(\chi)]}{2}=F_g+F_a+F_h=0\,.
 \end{eqnarray}
 The quantities  $F_g=-\frac{\alpha'[\tilde{\rho}+\tilde{p}]}{2}$, $F_a=\frac{2(\tilde{p}_t-\tilde{p}_r)}{\chi}$ and $F_h=-\frac{d\tilde{p}_r(\chi)}{d\chi}$ are the gravitational,  anisotropic and  hydrostatic forces, respectively. The behavior of the TOV equation of model (\ref{sol}) is shown in Fig. \ref{Fig:8} which
  indicates that the anisotropic and  gravitational  forces are negative and
 dominated by the  hydrostatics force  which is positive  and
keeps the system in static equilibrium in the case of TEGR whereas in the higher order torsion the  anisotropic and  hydrostatic  forces are positive and
 dominated by the  gravitational force   which is negative and keeps
 the system in static equilibrium.
\subsection{\bf The adiabatic index}
\begin{figure}
\centering
\subfigure[~The adiabatic index of  (\ref{sol}) when $\varepsilon=0$ ]{\label{fig:Ad0}\includegraphics[scale=0.3]{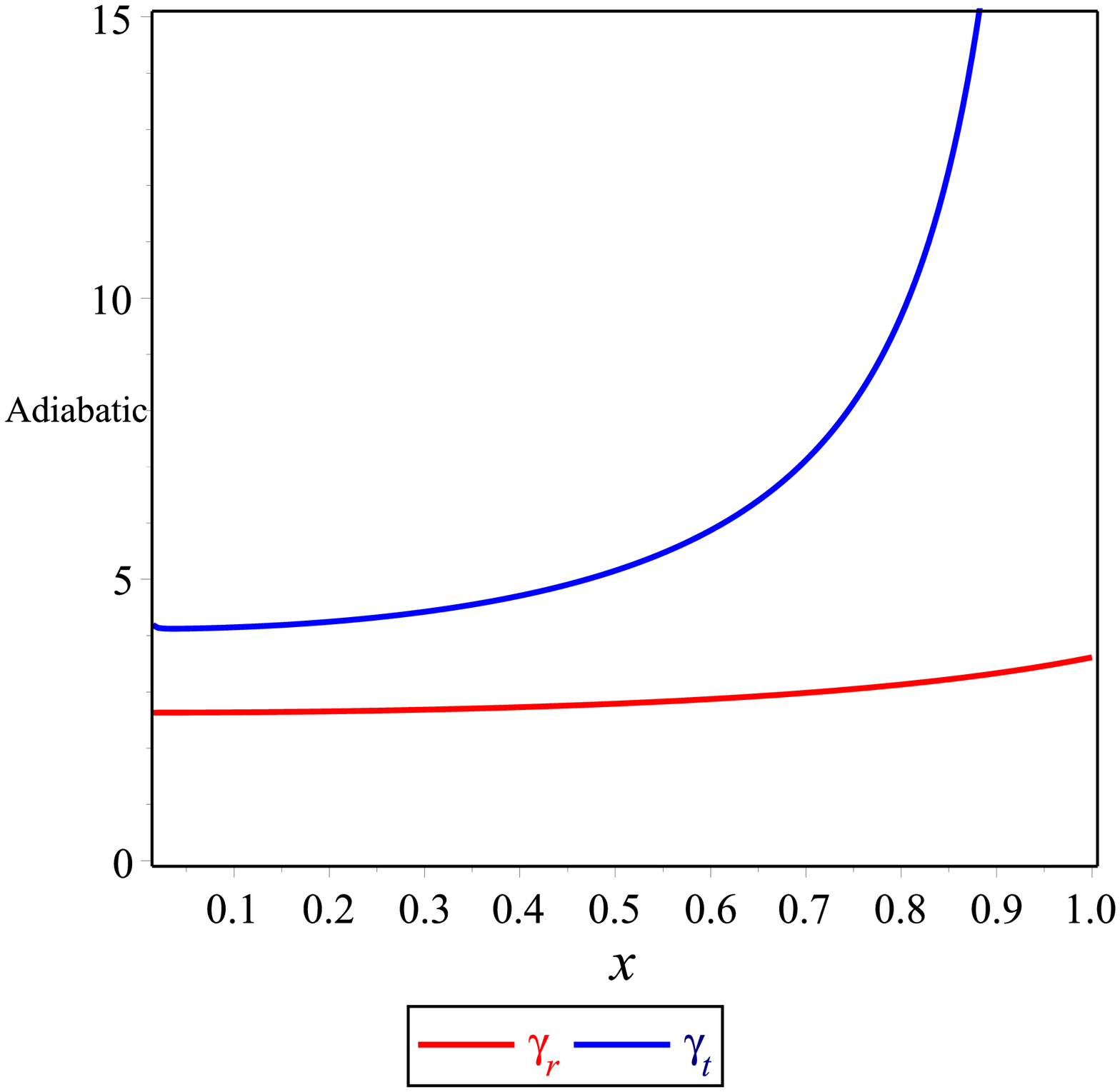}}
\subfigure[~The adiabatic index of  (\ref{sol}) when $\varepsilon=0.5$]{\label{fig:Ad}\includegraphics[scale=0.3]{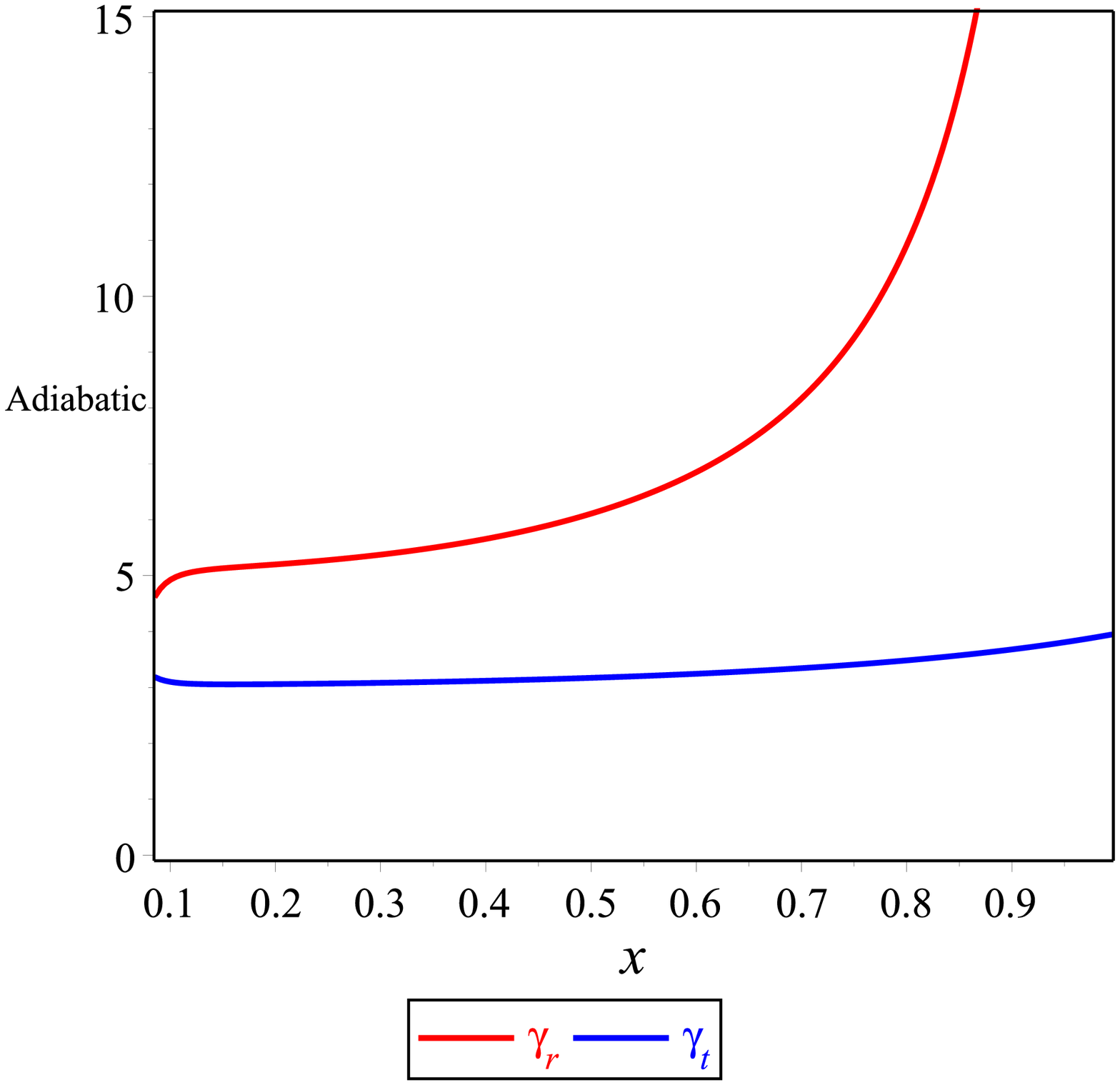}}
\caption[figtopcap]{\small{{Adiabatic index  of solution  (\ref{sol}) when $\varepsilon=0$ and $\varepsilon=0.5$}}}
\label{Fig:9}
\end{figure}

It is well-known that the adiabatic index is $\gamma$  and defined as \cite{1994MNRAS.267..637C,1997PhR...286...53H}
\begin{eqnarray}\label{aix} \gamma=\Big(1+\frac{\tilde{\rho}}{\tilde{p}}\Big)\frac{d\tilde{p}}{d\tilde{\varepsilon}}\,,
 \end{eqnarray}
which allows us to link the structure of a symmetrically spherical
static object and the EoS of the interior solution so that we can investigate its
 stability \cite{Moustakidis:2016ndw}. For any  interior solution to exhibit stability, it is important that  its adiabatic index be greater than $4/3$ \cite{1975A&A....38...51H} and if   $\gamma=\frac{4}{3}$  then  the isotropic
sphere will be in neutral equilibrium.  In accordance with the  work of Chan et al. \cite{10.1093/mnras/265.3.533},  the
condition of the stability of a relativistic anisotropic sphere $\gamma >\Gamma$ must be satisfied  such that \begin{eqnarray}\label{ai} \Gamma=\frac{4}{3}-\left\{\frac{4(\tilde{p}-\tilde{p}_t)}{3\lvert \tilde{p}'\lvert}\right\}_{max}\,.
 \end{eqnarray}
 The behavior of the adiabatic index is shown in Figure \ref{Fig:9} which ensures the  stability condition of (\ref{sol}).

\begin{table*}[t!]
\caption{\label{Table0}%
Values of model parameters when $\varepsilon=0.5$}
\begin{ruledtabular}
\begin{tabular*}{\textwidth}{lccccccc}
{{Pulsar}} & Mass ($M_{\odot}$) & {Radius (km)} & {$f_0$} & {$f_1$} &{$f_2$}& \\ \hline
&&&&&&&\\
J0437-4715 &$1.44^{+0.7}_{-0.07}$&$13.6^{+0.9}_{-0.8}$ \cite{2016MNRAS.455.1751R,2019MNRAS.490.5848G}&0.2454095942&-0.6438925458&0.39855633\\
&&&&&&&\\
J0030+0451 &$1.44^{+0.15}_{-0.16}$&$13.02^{+1.25}_{-1.06}$ \cite{2019ApJ...887L..24M}&0.2841372756&-0.7327677384&0.44874578\\
&&&&&&&\\
J0030+0451 &$1.34^{+0.15}_{-0.16}$&$12.71^{+1.14}_{-1.19}$\cite{2019ApJ...887L..21R}&0.2665730489&-0.6927871908&0.4263174699\\
&&&&&&&\\
EXO 1785 - 248 & $1.3^{+0.2}_{-0.2}$ & $8.849^{+0.45}_{-0.45}$ &0.003403660033&-0.01018791514&0.006784254813\\
&&&&&&&\\
Cen X-3 & $1.49^{+0.08}_{-0.08}$ & $9.178^{+0.13}_{-0.13}$ &0.5216844162&-1.227938771&0.7072044229 \\
&&&&&&&\\
RX J 1856 -37 & $0.9^{+0.2}_{-0.2}$ & $\simeq 6$ & 0.6103362288 & -1.393834058 &0.7865648502 \\
&&&&&&&\\
Her X-1 & $0.85^{+0.15}_{-0.15}$ & $8.1^{+0.41}_{-0.41}$ & 0.2892875212 &-0.7441973693 &0.4552206652  \\
&&&&&&\\

\end{tabular*}
\end{ruledtabular}
\end{table*}
\newpage
\begin{table*}[t!]
\caption{\label{Table1}%
Values of physical quantities when $\varepsilon=0$}
\begin{ruledtabular}
\begin{tabular*}{\textwidth}{lcccccccccc}
{{Pulsar}} &{$\tilde{\rho}\lvert_{_{_{\chi=0}}}$} & {$\tilde{\rho}\lvert_{_{_{\chi=1}}}$} & {$\frac{d\tilde{p}_r}{d\tilde{\rho}}\lvert_{_{_{\chi=0}}}$} & {$\frac{d\tilde{p}_r}{d\tilde{\rho}}\lvert_{_{_{\chi=1}}}$}
  & {$\frac{d\tilde{p}_t}{d\tilde{\rho}}\lvert_{_{_{\chi=0}}}$} & {$\frac{d\tilde{p}_t}{d\tilde{\rho}}\lvert_{_{_{\chi=1}}}$}&{$(\tilde{\rho}-\tilde{p}_r-2\tilde{p}_t)\lvert_{_{_{\chi=0}}}$}&{$(\tilde{\rho}-\tilde{p}_r-
  2\tilde{p}_t)\lvert_{_{_{\chi=1}}}$}&{$z\lvert_{_{_{\chi=1}}}$}&\\ \hline
&&&&&&&&&&\\ \hline
&&&&&Millisecond Pulsars&&& \\
&&&&& with White Dwarf &&& \\
&&&&&  Companions&&& \\
&&&&&&&&&&\\ \hline
J0437-4715&0.598&0.27$\times10^{-2}$ &0&-9$\times10^{-28}$ &-1&-1&0.461 & 0.81$\times10^{-2}$&0.039\\
&&&&&&&&&&\\

J0030+0451&0.67&0.3$\times10^{-2}$ &0 &-9$\times10^{-29}$&-1&-1&0.497&0.88$\times10^{-2}$&0.041\\
&&&&&&&&&&\\
J0030+0451&0.64&0.31$\times10^{-2}$ &0 &-9$\times10^{-26}$ &-1&-1&0.48 & 0.01&0.042\\
&&&&&&&&&&\\\hline
&&&&&Pulsars Presenting&&& \\
&&&&& Thermonuclear &&& \\
&&&&& Bursts &&& \\
&&&&&&&&&&\\ \hline
EXO 1785 - 248&1.1&0.0064&0 &-6$\times10^{-20}$ &-1&-1&0.58 &0.02&0.062\\
&&&&&&&&&&\\
Cen X-3 &1.06 &006 &0 &-6$\times10^{-22}$ &-1&01&0.58&0.018&0.069 \\
&&&&&&&&&&\\
RX J 1856 -37 &1.17&0.014&0 &-2.1$\times10^{-9}$&-1&-0.99&0.567&0.0417&0.095 \\
&&&&&&&&&&\\
Her X-1 &0.64&0.007 &0 &-2$\times10^{-10}$ &-1&-1&0.48&0.02& 0.07 \\
\end{tabular*}
\end{ruledtabular}
\end{table*}

\begin{table*}[t!]
\caption{\label{Table11}%
Values of physical quantities when $\varepsilon=0.5$}
\begin{ruledtabular}
\begin{tabular*}{\textwidth}{lcccccccccc}
{{Pulsar}} &{$\tilde{\rho}\lvert_{_{_{\chi=0}}}$} & {$\tilde{\rho}\lvert_{_{_{\chi=1}}}$} & {$\frac{d\tilde{p}_r}{d\tilde{\rho}}\lvert_{_{_{\chi=0}}}$} & {$\frac{d\tilde{p}_r}{d\tilde{\rho}}\lvert_{_{_{\chi=1}}}$}
  & {$\frac{d\tilde{p}_t}{d\tilde{\rho}}\lvert_{_{_{\chi=0}}}$} & {$\frac{d\tilde{p}_t}{d\tilde{\rho}}\lvert_{_{_{\chi=1}}}$}&{$(\tilde{\rho}-\tilde{p}_r-2\tilde{p}_t)\lvert_{_{_{\chi=0}}}$}&{$(\tilde{\rho}-\tilde{p}_r-
  2\tilde{p}_t)\lvert_{_{_{\chi=1}}}$}&{$z\lvert_{_{_{\chi=1}}}$}&\\ \hline
&&&&&&&&&&\\ \hline
&&&&&Millisecond Pulsars&&& \\
&&&&& with White Dwarf &&& \\
&&&&&  Companions&&& \\
&&&&&&&&&&\\ \hline
J0437-4715&1.2&0.9 &-1.14&0.28&-2.4$\times10^{-7}$&0.16&0.92& 0.86&0.77\\
&&&&&&&&&&\\

J0030+0451&1.3&0.97 &1.4 &0.29&0.32$\times10^{-5}$&0.17&0.98&0.93&0.97\\
&&&&&&&&&&\\
J0030+0451&1.3&0.94 &9.3 &0.28 &0.16$\times10^{-5}$&0.16&0.96 & 0.9&0.89\\
&&&&&&&&&&\\\hline
&&&&&Pulsars Presenting&&& \\
&&&&& Thermonuclear &&& \\
&&&&& Bursts &&& \\
&&&&&&&&&&\\ \hline
EXO 1785 - 248&1.7&1.2&0 &0.31 &3.9$\times10^{-7}$&0.19&1.1 &1.2&2.8\\
&&&&&&&&&&\\
Cen X-3 &1.8 &1.3 &0 &0.31&-0.15$\times10^{-3}$&0.2&1.1&1.2&4.4 \\
&&&&&&&&&&\\
RX J 1856 -37 &1.8&1.3&0 &0.31&1.7$\times10^{5}$&0.2&1.1&1.2&3.8 \\
&&&&&&&&&&\\
Her X-1 &1.4&1 &0 &0.3&0.1&0.2&1&0.95& 0.8 \\
\end{tabular*}
\end{ruledtabular}
\end{table*}

\section{Discussion and conclusion}\label{S7}

In the context of quadratic $f\mathcal{(T)}=\mathcal{T}+\epsilon \mathcal{T}^2$ modified gravity and for a KB metric potentials we gave a  precise interior solution that has the ability to describe real compact star configurations. The regularity conditions at the origin as well as at the surface of the stellar show that our solution has a good attitude over the stellar structure using the white dwarf companion $PRS J0437-4715$ . Moreover,  we show that the anisotropy of this model has a positive value when the dimensional parameter $\varepsilon\neq 0$ which can be interpreted as a repulsive force.  This fact  is because  the tangential pressure is greater than the radial pressure, i.e., $p_t>p_r $ \cite{2019JApA...40....8S}. When the dimensional parameter $\varepsilon=0$, which is the GR case, we got a negative anisotropy parameter.  Furthermore, we study the issue of stability and showed that the derived model is stable against the different forces (gravitational, hydrostatic, anisotropic, and electromagnetic) acting on it.  We also calculated the sound of speed and showed that it is consistent with realistic compact stars. Finally, we calculated the adiabatic index of our model and showed that it is also consistent with a realistic physical star. It is worth mention that the adiabatic index presented in \cite{Singh:2019ykp} has a negative value that is not consistent with realistic stellar models. The main reason for getting negative adiabatic index \cite{Singh:2019ykp} is due to the use of vanishing radial pressure. This indicates, in a clear way, that our assumption of the KB  metric potentials (\ref{eq:pot})  are the physical assumption that makes the resulting stellar model in the frame of quadratic $f(T)$ gravitational theory is consistent with real stellar objects.

{ In this study and for the first time we show the effect of the higher order torsion in modified teleparallel gravitational theories by assuming the form of $f\mathcal{(T)}=\mathcal{T}+\epsilon \mathcal{T}^2$.  We showed that the effect of the dimensional parameter $\epsilon$ made all the physical parameter consistent with a real stellar pulsar as we discussed in the context of this study. Moreover,  we tested our model over a  wide range of reported
observed values of masses and radii of pulsars are reported in (Tables I, II and III). We show  that the  fit is good also in these cases.}

The present approach can be summarized as follows: we have used a non-diagonal form of the tetrad field that gives a null value of the torsion tensor as soon as the metric potentials approach $1$. This is a crucial condition for any physical tetrad field as reported in various studies \cite{DeBenedictis:2016aze,Abbas:2015yma,Momeni:2016oai,2015Ap&SS.359...57A,Chanda:2019hyh,Debnath:2019yge,Ilijic:2018ulf}. Moreover, we stress the fact that the tetrad field (\ref{tet1}) is a necessary tetrad in the application of any spherically symmetrical interior/exterior solution in the frame of $f(T)$ because it makes the off-diagonal components of the field equations vanishing identically.

{ Bhatti et al. \cite{2018IJMPD..2750044B} have discussed the physical features of
compact stars using the ansatz of Krori and Barua in the frame  of $f(G,{\cal T})$, where $G$ is the Gauss-Bonnet term and ${\cal T}$ is the trace of the energy-momentum tensor, showing the behavior of material variables through plots and discussed the physical viability of compact stars through energy conditions.  Moreover, in the frame of $f(R)$, discussions of the behavior of different forces, equation of state parameter, measure of anisotropy and TOV equation in the modeling of stellar structures have been carried out. The comparison from their graphical representations provided clear evidence for the realistic and viable $f(R)$ gravity models at both theoretical and the astrophysical scale \cite{Yousaf:2017lto}. }

We can conclude that the comparison of our exact interior solution with the physical parameters of pulsars gives indications that the model can realistically represent observed systems.  Furthermore, the approach can be extended to a large class of metrics and anisotropy if the above physical requirements are satisfied. However, a detailed confrontation with observational data is needed.  This will be our forthcoming study.

\section*{Acknowledgement}
The work of KB was supported in part by the JSPS KAKENHI Grant Number JP21K03547.


%

\end{document}